\def\upd{{\rm d}}

\documentclass[nofootinbib,pre,showpacs]{revtex4}
\usepackage{graphicx}

\begin{document}

\title{Nonequilibrium statistical mechanics of shear flow: invariant quantities and current relations}

\author{A. Baule$^1$ and R. M. L. Evans$^2$}

\affiliation{$^1$The Rockefeller University, 1230 York Avenue, New York, NY 10065, USA\\
$^2$School of Physics and Astronomy, University of Leeds, Leeds LS2 9JT, United Kingdom}

\date{\today}

\begin{abstract}

In modeling nonequilibrium systems one usually starts with a definition of the microscopic dynamics, e.g., in terms of transition rates, and then derives the resulting macroscopic behavior. We address the inverse question for a class of steady state systems, namely complex fluids under continuous shear flow: how does an externally imposed shear current affect the microscopic dynamics of the fluid? The answer can be formulated in the form of \textit{invariant quantities}, exact relations for the transition rates in the nonequilibrium steady state, as discussed in a recent letter [A. Baule and R. M. L. Evans, Phys. Rev. Lett. 101, 240601 (2008)]. Here, we present a more pedagogical account of the invariant quantities and the theory underlying them, known as the nonequilibrium counterpart to detailed balance (NCDB). Furthermore, we investigate the relationship between the transition rates and the shear current in the steady state. We show that a fluctuation relation of the Gallavotti-Cohen type holds for systems satisfying NCDB.

\end{abstract}

\pacs{05.70.Ln, 05.20.Jj, 83.60.Rs}

\maketitle

\section{Introduction}

One of the greatest triumphs of nineteenth century physics was the development of equilibrium statistical mechanics --- a theory that exactly determines the probability of encountering any given arrangement of (the positions and momenta of) a large collection of interacting bodies (atoms, molecules, etc.), even though their equations of motion are intractable. From that probabilistic description, macroscopic properties \textit{en masse} can be found, such as a fluid's pressure, density, anisotropy, etc. The theory assumes the existence of thermodynamic equilibrium, an asymptotic long-time limit solution to the entire dynamics in which the statistical properties are unchanging and independent of initial conditions. Conversely, since the birth of statistical mechanics an exact prediction of the statistics of the particles' motion has been unobtainable in the presence of flow, instead requiring various approximations or simplifications, which are the bread and butter of much present-day research into nonequilibrium phenomena.

There is a certain class of highly nonequilibrium states of matter that possesses time-invariant statistics, and therefore might be approachable by exact statistical analysis without simplified modeling or near-equilibrium approximations. Such states arise when continuous shear flow is applied to complex fluids --- materials composed of classical particles much larger than atoms. Under flow the structure of complex fluids can be radically re-ordered, even to the point of undergoing shear-induced phase transitions.

For instance, a solution of amphiphiles in the worm-like micelle phase can exhibit a negative gradient in the characteristic flow curve if driven at high shear rates. This means that for a range of shear rates the viscosity decreases under shear. If the fluid is continuously sheared in this regime, it becomes mechanically unstable and separates into `bands' of high and low shear rate \cite{Spenley93}. Also, at higher concentration the solution of amphiphiles can undergo a phase transition from a lamellar phase into complex densely packed onion-like structures \cite{Roux93}. Both phenomena are structural phase transitions that are controlled by shear rate in addition to the usual parameters (temperature, pressure, concentration) and  are examples in which the fluid undergoes drastic microscopic re-configuration in response to the imposed (macroscopic) driving at the boundaries.

A fluid in a sheared steady state is described by the same Hamiltonian as at equilibrium, since no external field is applied to drive the system; only the boundary conditions are different. Nevertheless, since they contain non-zero fluxes, such states fall outside the jurisdiction of equilibrium statistical mechanics, despite exhibiting all the vast variety of reproducible behaviors, structures, and transitions seen in equilibrium thermodynamic systems. 

A statistical description of complex fluids under continuous shear flow, starting from first principles, is provided by a theory of nonequilibrium transition rates, known as the nonequilibrium counterpart to detailed balance (NCDB) \cite{EvansR04,EvansR05,Simha08}. This theory addresses the following question: given an externally imposed macroscopic shear current at the boundaries of the fluid, what is the effect on the microscopic dynamics (the transition rates) in the bulk of the fluid? In the absence of any driving, that is, at thermal equilibrium, the microscopic transition rates are constrained to satisfy the principle of detailed balance, namely that the ratio of forward to reverse transition rates between any pair of microstates must equal the Boltzmann factor of their energy difference:
\begin{eqnarray}
\label{canDB}
\frac{\omega_{ij}}{\omega_{ji}}=e^{-\beta(E_j-E_i)},
\end{eqnarray}
where $\omega_{ij}$ denotes the transition rate from microstate $i$ to $j$ and $\beta$ the inverse temperature. Detailed balance is a consequence of the influence of the fluid's thermal surroundings: it is immersed in a larger volume of fluid at a certain temperature which acts as an equilibrium heat reservoir. The stochastic influence of this heat reservoir puts constraints on the allowed microscopic transitions in the form of Eq.~(\ref{canDB}).

If shear is imposed at the boundaries of the heat reservoir detailed balance no longer holds. However, a fluid region within the bulk continues to receive stochastic forces from the reservoir, which is now itself under flow. In the steady state this nonequilibrium heat reservoir imposes constraints on the transition rates according to NCDB, which can be expressed in the form of a one-to-one mapping between the transition rates at equilibrium and those in the sheared steady state \cite{EvansR05}
\begin{eqnarray}
\label{NCDB_intro}
\Omega_{ij}(\nu)=\omega_{ij}\,e^{\nu \Delta x_{ji}+\Delta q_{ji}(\nu)}.
\end{eqnarray}
Here, $\Omega_{ij}(\nu)$ denotes a transition rate in the sheared steady state, parametrized by $\nu$, a Lagrange multiplier characterizing the driving strength of the reservoir. Under the influence of shear forces applied at the boundaries of the whole ensemble, the rates are thus enhanced or attenuated with respect to equilibrium. The factor $e^{\nu \Delta x_{ji}}$ exceeds unity if the transition $i\rightarrow j$ involves a conformational change that increments the shear strain by a positive amount $\Delta x_{ij}$. This factor simply boosts every transition in the forward direction irrespective of the state space structure. By itself, it would represent a simple mean-field expression for the driven transition rates. Important non-mean-field information about the global properties of the state space is contained in the quantities $\Delta q_{ji}(\nu)$, whose precise definition is given further below, in Sec.~\ref{Sec_NCDB}.

Recently, it has been demonstrated that the constraints of NCDB can be cast into the form of \textit{invariant quantities} that remain unchanged by the driving \cite{Baule08b} and apply to any pair of microstates in the following way: (i) The product of forward and reverse transition rates remains invariant under the driving, i.e., is the same in the equilibrium and in the sheared steady state: $\omega_{ij}\omega_{ji}=\Omega_{ij}\Omega_{ji}$ (where $\Omega_{ij}$ denotes the transition rate in the sheared state). (ii) The difference of total exit rates remains invariant: $\sum_k \left( \Omega_{ik}-\Omega_{jk}\right)=\sum_k \left(\omega_{ik}- \omega_{jk}\right)$. These invariant quantities represent exact relations for the transition rates in the sheared steady state, arbitrarily far away from equilibrium and represent a prediction of NCDB that can be tested in an experiment. Furthermore, using a graph representation of NCDB we have devised a systematic method to calculate the driven transition rates from a set of algebraic equations for arbitrary configurations of the system state space, thus greatly enhancing the applicability of the theory.

In this article we present a more pedagogical account of these recent results and the theory of NCDB underlying them. In particular, we provide a detailed derivation of the various representations of NCDB from a nonequilbrium sheared ensemble. Furthermore, we investigate the properties of the shear current in systems satisfying NCDB and show that the shear current exhibited by an individual system trajectory satisfies a fluctuation relation of the Gallavotti-Cohen type.

The remainder of this article is organized as follows. The nonequilibrium ensemble on which the derivation of NCDB relies is introduced in Sec.~\ref{Sec_ensemble}. In Sec.~\ref{Sec_NCDB} we present a detailed review of the different representations of NCDB as previously discussed in Refs.~\cite{EvansR04,EvansR05,Simha08}. A graph representation is presented in Sec.~\ref{Sec_Graph} providing an intuitive way to discuss Master equation systems in discrete state spaces. The invariant quantities are derived in Sec.~\ref{Sec_Q}, where we also formulate the systematic calculation method for the transition rates in the sheared steady state. The relationship between the shear current and the transition rates is discussed in Sec.~\ref{Sec_cur}. Our rather formal results are elucidated in two simple hopping models which allow for an explicit calculation of the driven transition rates and other relevant quantities of the NCDB formalism (Secs.~\ref{Sec_2} and \ref{Sec_3}). Even in these simple models it is evident that the predictive power of NCDB goes well beyond simple mean-field theories.

\section{Nonequilibrium ensemble and path entropy}
\label{Sec_ensemble}

In order to describe nonequilibrium states of complex systems subject to noise, one usually relies on a probabilistic description in terms of transition rates. In such an approach the main quantity of interest is the set of probability distributions $\{p_i(t)\}$ over states $i=1,...,n$, which express the probability to find the system in state $i$ at time $t$. The nature of these states depends on the level of description; for reasons that will become clear below they are considered to be classical microstates in the following. The dynamical evolution of the system is then governed by the Master equation, a balance equation for the probability:
\begin{eqnarray}
\label{master}
\frac{\upd}{\upd t}p_i(t)=\sum_{\{j\}}[\omega_{ji}\,p_j(t)-\omega_{ij}\,p_i(t)].
\end{eqnarray}
The sum is here taken over the set of states $\{j\}$ connected with $i$, where $\omega_{ij}$ denotes the rate of transition from state $i$ to state $j$. The difference $\omega_{ji}p_j(t)-\omega_{ij}p_i(t)$ is interpreted as the microscopic probability current between states $i$ and $j$, and therefore Eq.~(\ref{master}) states the conservation of probability. For a probabilistic interpretation of the $p_i(t)$ we require the conditions $0 \leq p_i(t) \leq 1$ and $\sum_{i=1}^n p_i(t)=1$.

A steady state is characterized by stationarity of the statistics, implying that all single-time distributions are time independent. If we set $\upd p_i(t)/\upd t=0$ in Eq.~(\ref{master}) we find that the condition for a steady state is given by the balance of total in- and outgoing flow for every state $i$
\begin{eqnarray}
\label{steady}
\sum_{\{j\}}[\omega_{ji}\,p_j-\omega_{ij}\,p_i]=0.
\end{eqnarray}
Nonequilibrium steady states are characterized by a non-zero net flow of particles, heat, etc., running through the system. By contrast, in equilibrium this flow vanishes, a fact which is manifest as the strong condition of detailed balance: the net probability current between \textit{any two configurations} of the system is zero at equilibrium. This is expressed as
\begin{eqnarray}
\label{DB}
\omega_{ji}\,p_j=\omega_{ij}\,p_i,
\end{eqnarray}
for all $i$, $j$ so that Eq.~(\ref{steady}) is trivially satisfied.

The condition of detailed balance is a physical property of transition rates in equilibrium systems and can be derived from Liouville's equation for closed, isolated, physical systems \cite{VanKampen}. `Physical' refers to the existence of an underlying Hamiltonian such that the dynamics is deterministic and microscopically reversible. `Isolation' implies that energy is a conserved quantity, i.e., a constant of motion. The trajectory in phase-space is then restricted to a shell of constant energy and no transitions between shells of different energy are possible. For a canonical system, these considerations apply in the same way to the combination system plus reservoir. The probability distribution of the system is then known to satisfy Boltzmann's law $p_i^{eq}\propto e^{-\beta E_i}$. In this context detailed balance can be interpreted as a statement of four fundamental properties characterizing system and heat reservoir: (i) ergodicity, (ii) microscopic reversibility, (iii) time-translation invariance of statistical properties, and (iv) conservation of energy. As a result of these properties there are $m/2$ constraints acting on $m$ transition rates of the system, expressed by Eq.~(\ref{canDB}).

\begin{figure}
\begin{center}
\includegraphics[width=10cm]{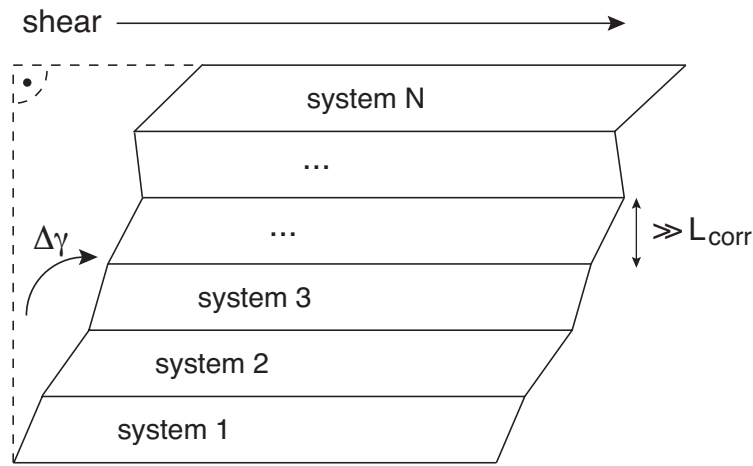}
\caption[The nonequilibrium shear ensemble.]{\label{Nfig_shear}A sketch of a large volume of sheared fluid: the fluid is hypothetically divided into a large number of subvolumes, the `systems'. The height of each subvolume is much larger than the correlation length $L_{corr}$ such that interactions with the surroundings, i.e., the `heat reservoir', only take place at the boundaries. The time evolution of a single system accumulates a certain amount of shear which is different for each system (indicated by the varying slopes). }
\end{center}
\end{figure}

Having specified an equilibrium state in this way, we may consider a particular class of driven steady state systems, namely fluids under continuous shear. Let us consider a fluid region far from the boundaries as our `system'. The heat reservoir consists of the fluid volume surrounding this region. If we assume that any correlation lengths are negligibly small compared with the fluid volume (a condition that may be unenforcable for turbulent flows), then this reservoir is only characterized by its macroscopic observables, which are in this case mean energy and mean shear rate. The stochastic influence on the system is here not that of an equilibrium reservoir, but of a reservoir which is itself in a nonequilibrium condition, under shear. In this case the properties (i)---(iii) remain valid. Ergodicity is generally difficult to prove rigorously even for the simplest systems at equilibrium, but it is assumed on empirical grounds, since experimental observations are repeatable, irrespective of precise initial conditions. Since the shear acts only at the boundaries of the reservoir, the dynamics of individual molecules is still governed by the same equations of motion as in equilibrium and thus microscopic reversibility holds. Property (iii) remains true by definition while (iv) continues to govern all interactions between the system and reservoir, or between different systems in an ensemble (Fig.~\ref{Nfig_shear}). We therefore find that the four conditions for detailed balance apply to a sheared fluid, amended by an additional conserved quantity, the total shear. These amended conditions give rise to a nonequilibrium counterpart to detailed balance, that can be derived from familiar statistical considerations in a straightforward way.

We continue, below, to derive the exact consequences of the above conditions, in terms of the resulting microscopic transition rates. That derivation is performed without approximation, and therefore applies arbitrarily far from equilibrium. However, as with any investigation, the degree to which a real experimental system is approximated by the idealized assumptions depends on the fluid in question. For instance, the applicability of any exact equilibrium calculation depends on how well a real system has been equilibrated and how well its symmetries are represented by the theory. For the driven cases considered here, the question of how close the experiment comes to a true steady state is even less straightforward. In an experiment on an isolated system and reservoir, the fluid is in principle only in a quasi-steady state because energy is continually pumped into the system by the driving force, tending to heat the fluid. Nevertheless, the above assumptions can be experimentally realized to arbitrarily high accuracy. Let us elucidate the types of fluid to which the above assumptions apply. 
In many complex fluids a solvent acts as a thermostat for the system of interest, as it possesses many more degrees of freedom than the mesoscopic particles of interest,
so the temperature remains relatively constant. In the limit of large ratio of solvent to complex degrees of freedom, the steady state condition becomes exactly realized. Since the relaxation time of the solvent is usually much smaller than that of the complex fluid, the solvent remains close to equilibrium even when the complex fluid is driven to a highly nonequilibrium state. In practice, experiments on complex fluids are able to obtain reproducible bulk steady-state behavior, indicating that the system is insensitive to spatial or temporal temperature gradients. These form the class of systems for which NCDB is expected to hold, and that reproducible behavior is the subject of our exact theory.

With these prerequisites in place, we can begin the derivation of NCDB from a nonequilibrium ensemble. Imagine a very large volume of fluid divided into a large number $N$ of sub-volumes or `systems'. These sub-volumes are still large enough that correlated regions within them are negligibly small (see Fig.~\ref{Nfig_shear}). The crucial idea in our derivation is to imagine this volume to be so large that it constitutes an ensemble in which the sub-volumes can be considered as representations of all possible realizations of our system under consideration. A certain constant shear rate $J$ is applied to the volume by moving top and bottom boundaries (while perpendicular dimensions are infinite or periodic). Each system $i$ then follows a particular trajectory in phase-space over a time period $\tau$ accumulating a certain amount of shear. In order to ensure steady state properties we will eventually take $\tau\rightarrow \infty$, so that any initial transient behavior has decayed.

With the nonequilibrium ensemble constructed in this way, the probability distribution of trajectories can be found by Gibbs's familiar method for deriving the probability distribution of a large collection of countable objects, which is exact when no correlations exist between those objects, as is the case here. The probability that a system follows phase-space trajectory $\Gamma$ is $p(\Gamma)=n_\Gamma/N$, where $n_\Gamma$ is the number of times $\Gamma$ is realized in the ensemble. Here, we take phase space and time to be discretized to make trajectories countable. Eventually the discrete intervals will vanish in the continuum limit and probabilities $p(\Gamma)$ will be replaced by distributions $p[\Gamma]\,{\cal D}\Gamma$.

The statistical weight of the ensemble $W_N$ is the number of distinct ways we can arrange the systems in the ensemble, which is given by the usual combinatorial formula
\begin{eqnarray}
W_N=\frac{N!}{\Sigma_\Gamma n_\Gamma!},
\end{eqnarray}
since systems following the same trajectory are indistinguishable. The most likely distribution of trajectories (the one adopted by the overwhelming majority of such ensembles) is the one with maximal statistical weight and can be found by maximizing the corresponding ensemble entropy $S_E=\ln W_N$. Applying Stirling's formula $\ln N!\approx N\ln N-N$ for large $N$ yields
\begin{eqnarray}
S_E=-N\sum_{\Gamma}p(\Gamma)\ln p(\Gamma).
\end{eqnarray}
Thus, the ensemble entropy per system is the \textit{path entropy}
\begin{eqnarray}
\label{Spath}
S_{\Gamma}=-\sum_{\Gamma}p(\Gamma)\ln p(\Gamma).
\end{eqnarray}
This path entropy is familiar from approaches to nonequilibrium statistical mechanics in the spirit of Jaynes's method of maximum entropy inference (MaxEnt). In fact, the first derivation of NCDB was based on this information theoretic approach \cite{EvansR04}. In the present framework of the nonequilibrium ensemble (cf. \cite{Simha08}) we are able to avoid certain controversial aspects of MaxEnt. Instead of a subjective interpretation of probabilities as representations of our state of knowledge, we have defined probabilities in terms of countable physical realizations. Furthermore, we have a clear idea of the prior set of trajectories that we reweight in the maximization procedure; it is the set of all physical trajectories with the same weights that they have at equilibrium. All unphysical ones are given zero weight a priori.

The distribution $p(\Gamma)$ is found by maximizing $S_\Gamma$ subject to the constraint
\begin{equation}
\label{shear_constraint}
  \sum_\Gamma p(\Gamma) \gamma(\Gamma) = J \tau
\end{equation}
due to conservation of shear. Here, $\gamma(\Gamma)$ is the total shear acquired by an individual system following path $\Gamma$. With $J=0$, the maximization just returns $p^{eq}(\Gamma)$, the equilibrium distribution of trajectories defined as the prior for the calculation. In the driven steady state we obtain instead the result
\begin{eqnarray}
\label{driven_p}
  p^{dr}(\Gamma)\propto \,p^{eq}(\Gamma)\, e^{\nu\gamma(\Gamma)},
\end{eqnarray}
where $\nu$ is the Lagrange multiplier associated with the shear constraint Eq.~(\ref{shear_constraint}).

This form of the nonequilibrium distribution is a direct consequence of the way in which the system is driven, namely by fluid at the boundaries such that the equations of motion of the driven system are the same as in equilibrium. In fact, the measure defined on trajectory space for the driven ensemble is the same as in equilibrium, the trajectories are only reweighted under the additional shear constraint giving rise to the factor $e^{\nu\gamma(\Gamma)}$.

Now that we are in possession of the statistical weight of nonequilibrium trajectories, we can consider the implications of Eq.~(\ref{driven_p}) on the individual transition rates, which give rise to all the complexity of the system's evolution. It is with this change of perspective that a useful and testable result can be obtained.

\section{Rules for transition rates}
\label{Sec_NCDB}

From now on we denote a transition rate between states $i$ and $j$ in the driven steady state with $\Omega_{ij}$ in order to distinguish it from the corresponding equilibrium transition rate $\omega_{ij}$ that satisfies equilibrium detailed balance. The transition rate is defined as the probability of making the transition $i\rightarrow j$ per unit time $\Delta t$, where it is understood that $\Delta t$ is so small that only one transition can occur. This means that transitions from $i$ to $j$ via a third state are neglected. In probability theory the definition of the transition probability $P(j|i)$ is usually in terms of a joint probability $P(j,i)$, i.e. $P(j|i)=P(j,i)/P(i)$. If the distribution of trajectories $p(\Gamma)$ is known, these probabilities of individual microstates are determined by counting those trajectories containing state $i$ at time $t$ and state $j$ at time $t+\Delta t$. A transition rate is then defined as $\Omega_{ij}=P(j|i)/\Delta t$ in the limit of vanishing $\Delta t$, where
\begin{eqnarray}
P(j|i) =\frac{\sum_{\Gamma | (i,j)\in\Gamma}p(\Gamma)}{\sum_{\Gamma | i\in\Gamma}p(\Gamma)}.
\end{eqnarray}
The sum $\sum_{\Gamma | (i,j)}$ denotes a summation over all trajectories $\Gamma$ containing the transition $i\rightarrow j$. Using Eq.~(\ref{driven_p}) a transition rate in the driven steady state can be written as
\begin{eqnarray}
\label{ratedelta}
\Omega_{ij}&=&\frac{P(j|i)}{\Delta t}=\frac{\sum_{\Gamma | (i,j)\in\Gamma}p^{eq}(\Gamma) e^{\nu\gamma(\Gamma)}}{\Delta t\sum_{\Gamma | i\in\Gamma}p^{eq}(\Gamma) e^{\nu\gamma(\Gamma)}}.
\end{eqnarray}
Inserting unity as the delta-function $\int\delta(\gamma-\gamma(\Gamma))\upd\gamma$ yields
\begin{eqnarray}
\Omega_{ij}=\frac{\int_{-\infty}^\infty e^{\nu\gamma}\sum_{\Gamma | (i,j)\in\Gamma}\delta(\gamma-\gamma(\Gamma))\,p^{eq}(\Gamma)\,\upd\gamma}{\Delta t \int_{-\infty}^\infty e^{\nu\gamma}\sum_{\Gamma | i\in\Gamma}\delta(\gamma-\gamma(\Gamma))\,p^{eq}(\Gamma)\,\upd\gamma}.
\end{eqnarray}
We notice that the average over the delta-function $\delta(\gamma-\gamma(\Gamma))$ is related to the conditional probability distribution $p_\tau^{eq}(\gamma|i,j)$:
\begin{eqnarray}
p_\tau^{eq}(\gamma|i,j)=\frac{\sum_{\Gamma | (i,j)\in\Gamma}\delta(\gamma-\gamma(\Gamma))p^{eq}(\Gamma)}{\sum_{\Gamma | (i,j)\in\Gamma}p^{eq}(\Gamma)},
\end{eqnarray}
The distribution $p_\tau^{eq}(\gamma|i,j)$ contains the probability that the system at equilibrium accumulates a total amount of shear $\gamma$ over a time period $\tau$ due to equilibrium fluctuations, given that it made a transition from $i$ to $j$. Similarly, the denominator in Eq.~(\ref{ratedelta}) leads to an expression containing $p_\tau^{eq}(\gamma|i)$. The subscript $\tau$ denotes the implicit dependence of these distributions on the duration of the trajectory $\Gamma$. The driven rate now reads
\begin{eqnarray}
\Omega_{ij}=\lim_{\tau\rightarrow\infty}\frac{\sum_{\Gamma | (i,j)\in\Gamma}p^{eq}(\Gamma)\int_{-\infty}^\infty p_\tau^{eq}(\gamma|i,j)\,e^{\nu\gamma}\,\upd\gamma}{\Delta t\sum_{\Gamma | i\in\Gamma}p^{eq}(\Gamma)\int_{-\infty}^\infty p_\tau^{eq}(\gamma|i)\,e^{\nu\gamma}\,\upd\gamma}.
\end{eqnarray}
Here, the $\tau\rightarrow\infty$ limit guarantees that the system has attained its stationary state. After factoring out the equilibrium rate
\begin{eqnarray}
\omega_{ij}=\frac{\sum_{\Gamma | (i,j)\in\Gamma}p^{eq}(\Gamma)}{\Delta t\sum_{\Gamma | i\in\Gamma}p^{eq}(\Gamma)},
\end{eqnarray}
we can write the driven transition rate as
\begin{eqnarray}
\label{drivenDB}
\Omega_{ij}=\omega_{ij}\;\lim_{\tau\rightarrow\infty}\frac{\int_{-\infty}^\infty p_\tau^{eq}(\gamma|i,j)\,e^{\nu\gamma}\,\upd\gamma}{\int_{-\infty}^\infty p_\tau^{eq}(\gamma|i)\,e^{\nu\gamma}\,\upd\gamma}.
\end{eqnarray}
We refer to this result as the \textit{canonical-flux} representation of NCDB for a transition rate $i\rightarrow j$ in the driven ensemble \cite{EvansR05}. Note that the driven rate is a function of the Lagrange multiplier $\nu$, the flux conjugate parameter. This parameter is the analogue of the inverse temperature $\beta$ which characterizes the heat transfer in canonical equilibrium systems. One notices that the driven transition rate is proportional to the equilibrium rate, enhanced or attenuated by a factor which is determined by the equilibrium statistics. If an equilibrium transition from $i$ to $j$ is likely to lead on to a certain amount of shear in the future, then this transition is enhanced in the driven steady state. As in the equilibrium case $m/2$ constraints act on $m$ driven transition rates, therefore Eq.~(\ref{drivenDB}) can be considered as a \textit{nonequilibrium counterpart to detailed balance}. The important conclusion is that for the class of driven steady states obeying the principles (i)---(iv) we obtain the microscopic dynamics of the nonequilibrium steady state from the corresponding equilibrium ensemble.

Depending on the physical system, many set of equilibrium (prior) rates $\omega_{ij}$ are possible, from exact and deterministic ones obtained from a Hamiltonian formulation to stochastic ones. Transitions which have zero probability in equilibrium (i.e., unphysical ones) are forbidden in the driven system as well.

\subsection{Alternative representations}
\label{Sec_NCDB_alt}

In the above derivation of NCDB the nonequilibrium ensemble was constrained by a fixed mean shear $J\tau$ (Eq.~(\ref{shear_constraint})). In analogy to equilibrium ensembles we can therefore consider Eq.~(\ref{drivenDB}) as a canonical representation of the driven ensemble: shear is a quantity which can be exchanged between the system and the nonequilibrium reservoir with a fixed average. Indeed both numerator and denominator in the enhancement factor take the form of an average where the conditional probability $p_\tau^{eq}(\gamma|i,j)$ (respectively $p_\tau^{eq}(\gamma|i)$) is weighted with the shear distribution $e^{\nu\gamma}$ over all possible shear $\gamma$. A \textit{microcanonical-flux} ensemble allows only for a fixed total shear $\gamma_0$. One can intuitively argue that in this ensemble the flux distribution has to be substituted by a Dirac delta function: $e^{\nu\gamma}\rightarrow\delta(\gamma-\gamma_0)$ (for an alternative derivation see \cite{EvansR05}). With this substitution Eq.~(\ref{drivenDB}) reads
\begin{eqnarray}
\label{drivenDB_micro}
\Omega_{ij}=\omega_{ij}\lim_{\tau\rightarrow\infty}\frac{ p_\tau^{eq}(\gamma_0|i,j)}{p_\tau^{eq}(\gamma_0|i)}.
\end{eqnarray}
This representation proves favorable when the conditional probability distributions of $\gamma_0$ are explicitly known. Similar to the ensembles in equilibrium statistical mechanics, microcanonical- and canonical-flux ensemble yield the same result in the thermodynamic limit, which is here the limit of large $N$ and $\tau$.

From Eq.~(\ref{drivenDB}) it is also possible to derive a $\tau$-independent representation of the driven transition rates \cite{EvansR05}. We reproduce the derivation here in full as it is used in the following investigation of NCDB in the context of discrete state spaces. Let us define the quantity $\Delta x_{ji}$ which denotes the shear contribution of the transition $i\rightarrow j$ in time $\Delta t$. Using $\Delta x_{ji}$ we can write the conditional probability distribution $p_\tau^{eq}(\gamma|i,j)$ as
\begin{eqnarray}
\label{app1}
p_\tau^{eq}(\gamma|i,j)=p_{\tau-\Delta t}^{eq}(\gamma-\Delta x_{ji}|j),
\end{eqnarray}
i.e., $p_\tau^{eq}(\gamma|i,j)$ is given by the probability distribution of accumulating the remaining shear $\gamma-\Delta x_{ji}$ in the remaining time $\tau-\Delta t$, starting from state $j$. Defining the function $m_i(\nu,\tau)$ as
\begin{eqnarray}
\label{app2}
m_i(\nu,\tau)\equiv \ln \int_{-\infty}^\infty p_\tau^{eq}(\gamma|i)\,e^{\nu\gamma}\,\upd\gamma,
\end{eqnarray}
allows us to rewrite Eq.~(\ref{drivenDB}) in the form
\begin{eqnarray}
\ln\frac{\Omega_{ij}}{\omega_{ij}}&=&\lim_{\tau\rightarrow\infty}\left(\ln\int_{-\infty}^\infty p_{\tau-\Delta t}^{eq}(\gamma-\Delta x_{ji}|i)\,e^{\nu\gamma}\,\upd\gamma-m_i(\nu,\tau)\right).
\end{eqnarray}
A change of the integration variable then leads to
\begin{eqnarray}
\ln\frac{\Omega_{ij}}{\omega_{ij}}&=&\nu\Delta x_{ji}+\lim_{\tau\rightarrow\infty}[m_j(\nu,\tau-\Delta t)-m_i(\nu,\tau)]\\
\label{drivenDB2}
&=&\nu\Delta x_{ji}+\lim_{\tau\rightarrow\infty}[m_j(\nu,\tau)-m_i(\nu,\tau)]-\zeta(\nu,\Delta t),
\end{eqnarray}
where
\begin{eqnarray}
\zeta(\nu,\Delta t)\equiv\lim_{\tau\rightarrow\infty}[m_j(\nu,\tau)-m_j(\nu,\tau-\Delta t)].
\end{eqnarray}
An important property of the function $\zeta$ is its state-independence, which follows from Eq.~(\ref{drivenDB2}) upon changing $\tau\rightarrow \tau+\Delta t$. To first order in $\Delta t$ we then have
\begin{eqnarray}
\zeta(\nu,\Delta t)=\lim_{\tau\rightarrow\infty}\frac{\partial}{\partial \tau} m_j(\nu,\tau)\Delta t.
\end{eqnarray}
We now introduce two important $\tau$-independent quantities \cite{EvansR05}. Firstly, we identify
\begin{eqnarray}
\label{longtau1}
\Delta q_{ji}=q_j(\nu)-q_i(\nu)&=&\lim_{\tau\rightarrow\infty}[m_j(\nu,\tau)-m_i(\nu,\tau)],
\end{eqnarray}
and secondly, the state-independent rate of change of $m_i(\nu,\tau)$ in the long-time limit is denoted as
\begin{eqnarray}
\label{longtau2}
Q(\nu)=\lim_{\tau\rightarrow\infty}\frac{\partial}{\partial \tau}m_i(\nu,\tau).
\end{eqnarray}
We therefore see that Eq.~(\ref{drivenDB2}) gives rise to
\begin{eqnarray}
\label{DDB}
\Omega_{ij}(\nu)=\omega_{ij}\,e^{\nu \Delta x_{ji}+\Delta q_{ji}(\nu)-Q(\nu)\Delta t},
\end{eqnarray}
which is a $\tau$-independent representation of NCDB. Equation~(\ref{DDB}) reduces to Eq.~(\ref{NCDB_intro}) in the continuous time limit, $\Delta t\to0$.

If we compare Eq.~(\ref{DDB}) with Eq.~(\ref{drivenDB}) we realize that the ratio of equilibrium Green's functions is here translated into three distinct factors. The factor $e^{\nu \Delta x_{ji}}$ measures the direct flux contribution of a transition and is large if either the flux carried by the transition $i\rightarrow j$ is large ($\Delta x_{ji}\gg 1$) or the system is strongly driven ($\nu \gg 1$). By itself this factor would simply boost every transition in the flux direction irrespective of the state space structure. The important extension to such mean-field ideas is expressed in the factor $e^{\Delta q_{ji}(\nu)}$, where $\Delta q_{ji}$ is formally defined as (cf. Eqs.~(\ref{app2}) and (\ref{longtau1}))
\begin{eqnarray}
\label{q_def}
\Delta q_{ji}(\nu)\equiv\lim_{\tau\rightarrow\infty}\left[\ln\frac{\int_{-\infty}^\infty p_\tau^{eq}(\gamma | j)\,e^{\nu\gamma}\,\upd\gamma}{\int_{-\infty}^\infty p_\tau^{eq}(\gamma | i)\,e^{\nu\gamma}\,\upd\gamma}\right].
\end{eqnarray}
This quantity measures the increase (or decrease) in probability that the system will go on to exhibit the imposed shear $\gamma$ if it performs the transition $i\rightarrow j$. Thus the rate of a transition not only depends on the immediate flux contribution, but also on the prospect for future flux.

Formally, the function $Q(\nu)$ can be defined as the scaled cumulant generating function of the shear current $J=\gamma/\tau$
\begin{eqnarray}
\label{Q_dev}
Q(\nu)\equiv\lim_{\tau\rightarrow\infty}\frac{1}{\tau}\ln\int_{-\infty}^\infty p_\tau^{eq}(J|i)\,e^{\nu\tau J}\,\upd J,
\end{eqnarray}
which agrees with Eqs.~(\ref{app2}) and (\ref{longtau2}) upon transforming from the (extensive) shear $\gamma$ to the (intensive) shear current $J$. The distribution $p_\tau^{eq}(J|i)$ contains the probability to observe a shear current $J$ over time $\tau$ in the equilibrium system, given it was initially in state $i$. Eq.~(\ref{Q_dev}) expresses the fact that $\nu$ and $J$ are conjugate quantities in the sense of large deviation theory and implies that $Q(\nu)$ is the Legendre transform of the rate function of $J$ \cite{EvansR05,Touchette09b}. Consequently, $Q(\nu)$ and $J$ are related via
\begin{eqnarray}
\label{Qflux}
\frac{\upd}{\upd \nu}Q(\nu)=J.
\end{eqnarray}
In the following we refer to the function $Q(\nu)$ as `flux potential' due to the analogy with the usual thermodynamic potentials.

In order to fully determine the transition rates of a system in the driven steady state according to Eq.~(\ref{DDB}) one therefore has to know the set of equilibrium transition rates, their individual shear contribution $\Delta x_{ji}$ as well as the flux potential $Q$ and the functions $\Delta q_{ji}$. At equilibrium the $\omega_{ij}$ are usually strongly constrained by detailed balance and symmetry considerations. The $\Delta x_{ji}$ are local and constant properties of the states. The crucial and non-trivial task is to find the potential $Q$ and the set of $\Delta q_{ji}$'s, which depend on the global structure of the state space.

\section{Graph representation for NCDB}
\label{Sec_Graph}

Master equation systems can be discussed intuitively with the help of a graph representation, in which vertices are assigned to the different states $i$ of the system and edges to the possible transitions \cite{Schnakenberg76}. If a transition is physically allowed to take place, i.e., $\omega_{ij}>0$, then equilibrium detailed balance demands that the reverse transition $\omega_{ji}$ is also non-zero. Only connected graphs are considered in order to satisfy the requirement of ergodicity. For the following discussion of driven steady states further assumptions are necessary. In order to guarantee that the system can exhibit a constant macroscopic steady state shear current $J$, it is assumed that the state space has a periodic structure along a direction $x$, which is associated with the amount of shear accumulated. This assumption is made without loss of generality since one period may be arbitrarily large. A transition $i\rightarrow j$ contributes the shear increment $\Delta x_{ji}\equiv x_j-x_i$. The periodic structure implies that, for a given period $n$, the $j+n$th vertex is of the same type as the $j$th with $x$-position shifted by a fixed amount. Likewise, if vertices $i$ and $j$ with $1\leq i,j\leq n$ are connected by an edge so are vertices $n+i$ and $n+j$.

\begin{figure}
\begin{center}
\includegraphics[width=8cm]{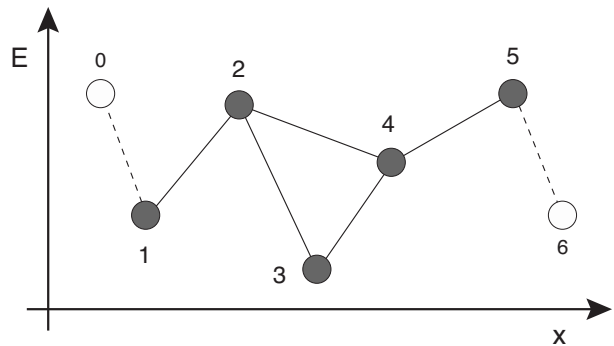}
\caption[Example diagram of a basic graph]{\label{Fig_Graph}Example diagram of a basic graph for a five-state system. The dotted line denotes an exterior edge connecting the same type of states in different periods. Here states $0$ and $5$, as well as states $1$ and $6$ are of the same type.}
\end{center}
\end{figure} 

We define the \textit{basic graph} as the graph corresponding to the non-periodic connected set of $n$ vertices. Its set of edges will be denoted interior edges in order to distinguish them from exterior edges connecting vertices of the basic graph with vertices of the next or preceding period. More precisely, an edge is exterior if it connects a vertex $i$ of the basic graph with a vertex $l$ where $l\geq n+1$ or $l \leq 0$. Obviously, exterior edges always occur in pairs connecting the same states in different periods. Using this convention one has to bear in mind that two states of the same type can be connected by more than one edge, namely by one interior edge and multiple exterior edges (connecting, e.g., to the previous or next period, or both). See Fig.~(\ref{Fig_Graph}) for a depiction of such a basic graph. The total number of transition rates in the system is $\sum_{i=1}^{n}d_i$, where $d_i$ is the degree (or connectivity) of the $i$th vertex of the simple graph including exterior edges. The minimal number of transition rates in an $n$-state driven system is $2n$ which corresponds to a graph in the form of a simple connected path. For this class of state spaces the problem of finding the driven transition rates has a particularly straightforward solution (see Sec.~\ref{Sec_net} below).

\begin{figure}
\begin{center}
\includegraphics[height=4cm]{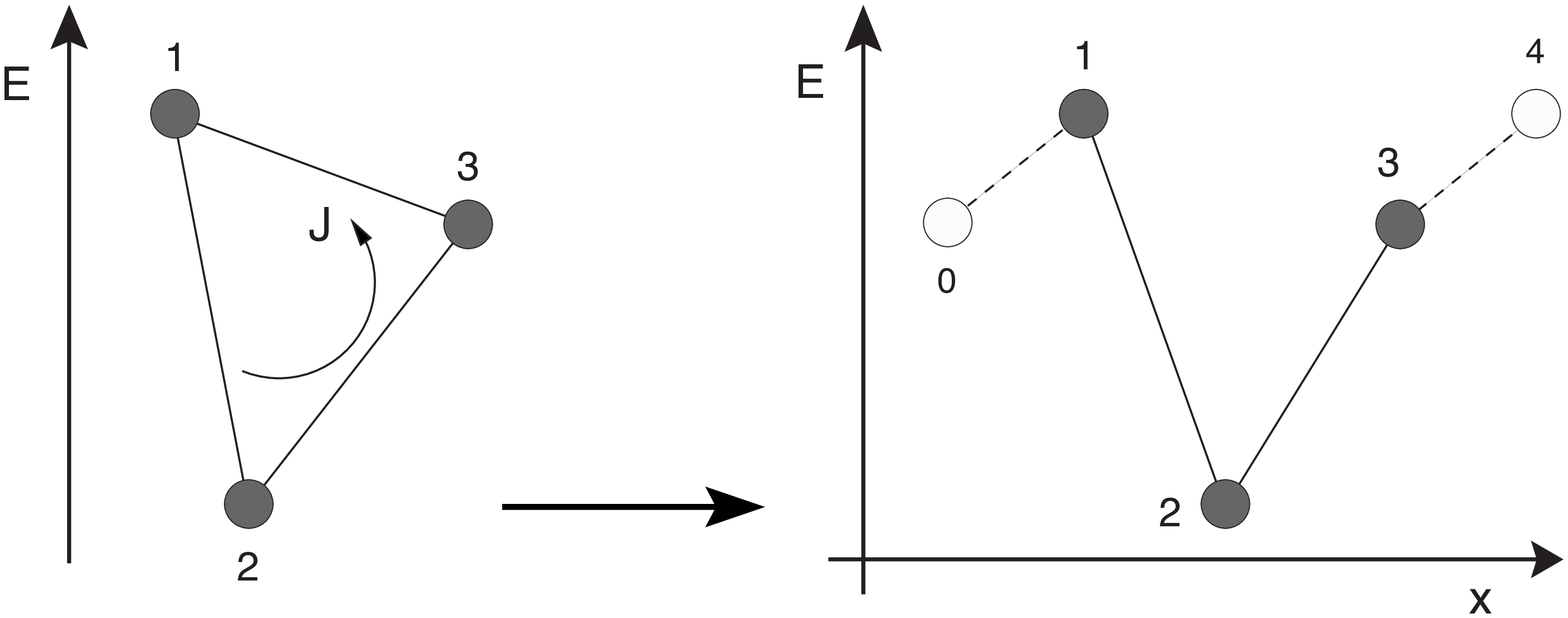}
\caption{\label{Fig_3ex}Example diagram of a three state system mapping onto a periodic graph structure.}
\end{center}
\end{figure}

This notion of a periodic graph structure is basically a convenient way to visualize the current in the state space. Systems with a limited number of states would usually be depicted as a basic graph alone without any exterior edges. Yet, if the system exhibits a nonequilibrium steady state with some kind of current, the periodicity automatically arises as a consequence of the flux carrying transitions. Consider for example the three state system in Fig.~\ref{Fig_3ex}. The only way the system can be in a driven steady state is by featuring a rotational current which is measured by the windings performed in time $\tau$. This then naturally maps onto a periodic network with a simple connected path as basic graph. The loop $1\rightarrow 2 \rightarrow 3\rightarrow 1$ in the original depiction then becomes an external loop $1\rightarrow 2 \rightarrow 3\rightarrow 4$, where state $4$ is of the same type as $1$ yet distinct due to the accumulated integrated current of one period. 

At this point it is appropriate to briefly discuss the distinction between external and internal loops, which is connected to Kolmogorov's criterion, an equivalent statement of equilibrium detailed balance \cite{EvansM02,Kelly}. In order to elucidate this we focus on transition rates in continuous time with $\Delta t\rightarrow 0$ such that Eq.~(\ref{DDB}) assumes the simpler form of Eq.~(\ref{NCDB_intro}).

If we consider a closed internal loop in the basic graph, i.e., a closed path leading back to the identical state such as, e.g., the loop $2\rightarrow 3\rightarrow 4\rightarrow 2$ in Fig.~(\ref{Fig_Graph}), we see that the product of transition rates following this loop in a given direction of rotation is the same as in equilibrium
\begin{eqnarray}
\Omega_{23}\Omega_{34}\Omega_{42}=\omega_{23}\omega_{34}\omega_{42}.
\end{eqnarray}
This is a simple consequence of the fact that both $\Delta x_{ji}$ and $\Delta q_{ji}$ are given as differences of state properties, implying that along a closed internal loop the product of exponential factors vanishes. Therefore, for any internal closed loop in the basic graph we have
\begin{eqnarray}
\label{kolm}
\Omega_{12}\cdots\Omega_{n-1,n}\Omega_{n1}=\omega_{12}\cdots\omega_{n-1,n}\omega_{n1}.
\end{eqnarray}
Since the equilibrium rates satisfy detailed balance, the ratio of forward and backward transitions is given as $\omega_{12}/\omega_{21}=e^{-\beta(E_2-E_1)}$. It is then easy to see that the following equality holds
\begin{eqnarray}
\label{kolm_eq}
\frac{\omega_{12}\cdots\omega_{n-1,n}\omega_{n1}}{\omega_{21}\cdots\omega_{n,n-1}\omega_{1n}}=1.
\end{eqnarray}
Due to Eq.~(\ref{kolm}) the same relation is true for the product of driven transition rates
\begin{eqnarray}
\label{kolm2}
\Omega_{12}\cdots\Omega_{n-1,n}\Omega_{n1}=\Omega_{21}\cdots\Omega_{n,n-1}\Omega_{1n}.
\end{eqnarray}
Kolmogorov's criterion now states that equilibrium detailed balance holds if and only if Eq.~(\ref{kolm2}) is satisfied for every closed path in state space. NCDB according to Eq.~(\ref{NCDB_intro}) satisfies this criterion for every internal loop. However, this does not lead to a contradiction, since Kolmogorovs criterion is violated for external loops in state space. For an external loop we have  instead of Eq.~(\ref{kolm2})
\begin{eqnarray}
\frac{\Omega_{12}\cdots\Omega_{n-1,n}\Omega_{n1'}}{\Omega_{21}\cdots\Omega_{n,n-1}\Omega_{1'n}}=e^{2\nu \Delta x_{1'1}},
\end{eqnarray}
where $1'$ denotes the state of type $1$ in the next period and $\Delta x_{1'1}$ is the accumulated shear or `length' of a period. The conclusion of this discussion is that the graph representation for NCDB in terms of a periodic graph structure is consistent if the system does exhibit a steady state current. In this case NCDB leads to a violation of Kolmogorov's criterion as expected.

At equilibrium the validity of detailed balance, or Kolmogorov's criterion respectively, implies that for every two states $k$ and $l$ the ratio
\begin{eqnarray}
S_{kl}\equiv\frac{\omega_{k1}\omega_{12}\cdots\omega_{n-1,n}\omega_{nl}}{\omega_{1k}\omega_{21}\cdots\omega_{n,n-1}\omega_{ln}} 
\end{eqnarray}
is independent of the path between $k$ and $l$. Out of equilibrium this path-independence is not generally expected. Instead one can consider $\ln S_{kl}$ as an `action functional' associated with a particular path and derive a fluctuation relation for the entropy production \cite{Lebowitz99}.

\section{Total exit rate relation}
\label{Sec_Q}

In the remainder of this chapter we focus on NCDB in the context of continuous time Markov chains, where NCDB is expressed in the form of Eq.~(\ref{NCDB_intro}). In this case it is possible to derive a fundamental relationship between the flux potential $Q(\nu)$ and the driven transition rates which is a central result of this article and leads to a variety of important conclusions for NCDB.

Consider an individual state $i$ connected to $d_i$ other states. The system in state $i$ spends a random time until it makes a transition to one of the connected states. For a continuous time Markov chain this waiting time has the exponential distribution \cite{Kelly}:
\begin{eqnarray}
\label{waitingt}
  h_i(t)=\sigma_i\,e^{-\sigma_i t},
\end{eqnarray}
where the total exit rate is defined as $\sigma_i\equiv\sum_{\{j\}}\omega_{ij}$. The probability that the particle jumps to site $j$ is then $P_{ij}=\omega_{ij}/\sigma_i$. Our quantities of interest are the conditional probabilities or Green's functions $p^{eq}_\tau(\gamma|i)$ which, if known, would fully specify the driven transition rates via Eqs.~(\ref{q_def}) and (\ref{NCDB_intro}). The Green's function for state $i$ can be determined by the following considerations. From state $i$ the system can only perform a transition to a connected state $j$ within the network, from where its further displacement is determined by the Green's function of that state $j$. Taking into account the waiting time in state $i$ and the probability $P_{ij}$ to perform the jump to state $j$, $p^{eq}_\tau(\gamma|i)$ is therefore related to the $d_i$ Green's functions of the neighbouring sites according to
\begin{eqnarray}
\label{greens}
p^{eq}_\tau(\gamma|i)=\int_0^\tau \upd t\, h_i(\tau-t)\sum_{\{j\}}P_{ij}\,p^{eq}_t(\gamma-\Delta x_{ji}|j)+\psi_i(\tau)\delta(\gamma),
\end{eqnarray}
where $\psi_i(\tau)\upd \tau$ denotes the probability that no jump has occurred\footnote{The contribution of $\psi_i(\tau)\delta(\gamma)$ to the Green's function $p^{eq}_\tau(\gamma|i)$ was erroneously neglected in the derivation presented in Ref.~\cite{Baule08b}. Nevertheless, the final result Eq.~(\ref{Qpi}) remains unchanged.} out of state $i$ up to time $\tau$: $\psi_i(\tau)=1-\int_0^\tau h_i(t)\upd t=e^{-\sigma_i \tau}$. In the next step we introduce the quantities $m_i(\nu,\tau)$, Eq.~(\ref{app2}), which are ultimately related to the functions $q_i(\nu)$ in the long time limit via Eq.~(\ref{longtau1}). Multiplying Eq.~(\ref{greens}) by $e^{\nu \gamma}$ and summing over all possible shear $\gamma$ from state $i$ yields, after a shift in the summation variable $\gamma$,
\begin{eqnarray}
e^{m_i(\nu,\tau)}=\int_0^\tau \upd t\,h_i(\tau-t)\sum_{\{j\}}P_{ij}\,e^{m_j(\nu,t)+\nu\Delta x_{ji}}+e^{-\sigma_i \tau}.
\end{eqnarray}
In the next step we can substitute $P_{ij}$, and $h_i(t)$ (Eq.~(\ref{waitingt})). Rearranging terms then yields
\begin{eqnarray}
e^{m_i(\nu,\tau)+\sigma_i\tau}=\int_0^\tau \upd t\,e^{\sigma_i t}\sum_{\{j\}}\omega_{ij}\,e^{m_j(\nu,t)+\nu\Delta x_{ji}}+1.
\end{eqnarray}
The integral can be removed by taking a derivative with respect to $\tau$ on both sides. This leads to
\begin{eqnarray}
\left(\frac{\partial}{\partial \tau}m_i(\nu,\tau)+\sigma_i\right)e^{m_i(\nu,\tau)+\sigma_i\tau}=e^{\sigma_i\tau}\sum_{\{j\}}\omega_{ij}\,e^{m_j(\nu,\tau)+\nu\Delta x_{ji}}.
\end{eqnarray}
Or, after further rearrangement,
\begin{eqnarray}
\label{Gmtau}
\frac{\partial}{\partial \tau}m_i(\nu,\tau)+\sigma_i&=&\sum_{\{j\}}\omega_{ij}\,e^{m_j(\nu,\tau)-m_i(\nu,\tau)+\nu\Delta x_{ji}}.
\end{eqnarray}
In the long time limit Eqs.~(\ref{longtau1}) and (\ref{longtau2}) hold. The result of taking the $\tau\rightarrow\infty$ limit in Eq.~(\ref{Gmtau}) is thus
\begin{eqnarray}
Q(\nu)+\sigma_i=\sum_{j}\omega_{ij}\,e^{\Delta q_{ji}(\nu)+\nu\Delta x_{ji}}.
\end{eqnarray}
On the right-hand side we can identify the transition rates in the driven steady state according to Eq.~(\ref{NCDB_intro}). We therefore obtain a fundamental relationship between the equilibrium transition rates, the corresponding rates in the driven steady state, and the flux potential $Q(\nu)$:
\begin{eqnarray}
\label{Qpi}
Q(\nu)=\Sigma_i(\nu)-\sigma_i,
\end{eqnarray}
where $\Sigma_i(\nu)\equiv\sum_{\{j\}}\Omega_{ij}(\nu)$. Eq.~(\ref{Qpi}) states that, for every state $i$, the total exit rate in the driven steady state differs from its equilibrium counterpart only by a flux dependent (but state independent) constant. On the basis of this central result a number of important implications of NCDB can be derived. It turns out that it is not necessary to know the Green's functions of the equilibrium system in order to determine the $\Delta q$'s and the driven transition rates. Rather, as we will see more explicitly below, the quantities of the NCDB formalism are intrinsically related to the graph structure via Eq.~(\ref{Qpi}).

\subsection{Invariant quantities}

It is now straightforward to formulate two sets of \textit{invariant quantities} for the sheared steady state. The first was found in Ref.~\cite{EvansR05} and is a consequence of the asymmetric property of $\Delta q_{ji}$ and $\Delta x_{ji}$. From Eq.~(\ref{NCDB_intro}), the\textit{`product constraint'} follows directly,
\begin{eqnarray}
\label{pcons}
\Omega_{ij}\Omega_{ji}=\omega_{ij}\omega_{ji}.
\end{eqnarray}
Secondly, Eq.~(\ref{Qpi}) directly implies the \textit{`total exit rate constraint'},
\begin{eqnarray}
\label{qcons}
\Sigma_i -\Sigma_j=\sigma_i-\sigma_j.
\end{eqnarray}
We therefore find that in the driven steady state both the product of forward and reverse transition rates and the difference of total exit rates for every pair of microstates are the same as in equilibrium and therefore invariant with respect to the driving. No near-equilibrium assumptions have been made in the derivation, so the above relations are both exact and valid arbitrarily far from equilibrium.

With the formalism devised here, the task of finding the microscopic nonequilibrium dynamics given an imposed macroscopic current is greatly simplified and follows straightforward rules. The above relations are furthermore accessible to verification in an experiment or simulation \cite{nextPRL}.

\subsection{Network rules}
\label{Sec_net}

With these results we are able to devise a systematic method to determine the driven transition rates for arbitrary graph configurations. Considering the whole basic graph with $n$ states, there are $n$ equations in the form of Eq.~(\ref{Qpi}). This set of equations is sufficient to determine all the unknown quantities, i.e., the flux potential $Q$ and the $\Delta q$'s, as we now show. Due to the relationship $\Delta q_{ij}=-\Delta q_{ji}$, every edge of the basic graph is associated to two transition rates (forward and backward transitions) depending on one $\Delta q$. The number of independent $\Delta q$'s is further constrained by closed paths (loops) in the graph, i.e., paths that begin and end at the same type of vertex, because the sum of $\Delta q$'s along such a path is obviously zero (loop constraint). One can then easily see that the total number of independent $\Delta q$'s in the basic graph is always $n-1$, as follows. Consider first the most simple basic graph configuration, namely all $n$ states connected as a simple path without any loops. In this case the number of edges is trivially $n$. Since one loop constraint is generated by the periodicity, there are $n-1$ independent $\Delta q$'s. From this simple connected graph all graphs of higher degrees can be generated by adding new edges. But adding an edge generates a new $\Delta q$ and at the same time a new loop constraint, so that the number of independent $\Delta q$'s always remains $n-1$.

We formulate the following \textit{network rules} for the calculation of the driven transition rates in networks of arbitrary connectivity:
\begin{itemize}
\item{\textit{Edge rule}. Every interior edge and every pair of exterior edges in the basic graph corresponds to two rates containing the dependence on one $\Delta q$. The driven transition rates are given by Eq.~(\ref{NCDB_intro}).}
\item{\textit{Vertex rule}. For every vertex in the basic graph the difference between the driven and equilibrium total exit rates equals the flux potential $Q$ (Eq.~(\ref{Qpi})).}
\item{\textit{Loop rule}. For every closed path of edges the sum of the $\Delta q$'s along this path is zero.}
\end{itemize}
In this formulation there are in total $n$ equations and $n$ unknowns, namely one $Q$ and $n-1$ $\Delta q$'s. The number of independent equations can always be further reduced by eliminating $Q$, such that one is essentially left with $n-1$ equations for $n-1$ unknown $\Delta q$'s. The solution of this system of equations fully specifies all the driven rates in the system as well as the flux $J$ which is related to the flux potential via $\upd Q(\nu)/\upd\nu=J$, Eq.~(\ref{Qflux}).
In the given framework the driven transition rates and the flux are determined as functions of the flux conjugated parameter $\nu$. 

An alternative, $\nu$-independent representation is based on Eqs.~(\ref{pcons}) and (\ref{qcons}). However, the number of constraints is here not sufficient to determine all the driven rates for general network structures. On the one hand there are $\sum_{i=1}^{n}d_i/2$ product constraints and $n-1$ exit rate constraints. On the other hand, for an arbitrary graph configuration, there are $\sum_{i=1}^{n}d_i$ transition rates. Therefore only for graphs with the topology of a simple connected path (where $d_i=2$), we can determine the rates completely from the invariant quantities without using the network rules. In this case we have $2n$ transition rates and $2n-1$ constraints stemming from the exact relations. The transition rates are fully determined if additionally the relationship between the transition rates and the current is provided. In this formulation the driven rates depend on $J$ directly instead of being parametrized in terms of the parameter $\nu$. The relationship between current and rates is further elucidated in the next section.

\section{Current relations}
\label{Sec_cur}

In Master equation systems the stationary current $J$ is generally defined as the average of the flux contributions of every state $i$ over the steady state distributions $p_i$
\begin{eqnarray}
\label{flux_def}
J=\sum_{i}p_i\sum_{\{j\}}\Delta x_{ji}\Omega_{ij}.
\end{eqnarray}
At equilibrium $J$ is identically zero due to detailed balance, Eq.~(\ref{DB}), and the property $\Delta x_{ji}=-\Delta x_{ij}$. In general, at steady state, the distribution $p_i$ is the solution of the Master equation (\ref{master}) under the condition of stationarity $\upd p_i(t)/\upd t=0$ and normalization $\sum_i p_i=1$. The solution can be formally obtained by matrix inversion as follows \cite{EvansM02}. In matrix notation Eq.~(\ref{master}) is given as (writing the transition rates now capitalized):
\begin{eqnarray}
\left(\begin{array}{ccccc}
-\Sigma_1 & \bar{\Omega}_{21} & \bar{\Omega}_{31} & \cdots & \bar{\Omega}_{n1}\\
\bar{\Omega}_{12} & -\Sigma_2 & \bar{\Omega}_{32} & \cdots & \bar{\Omega}_{n2}\\
\bar{\Omega}_{13} & \bar{\Omega}_{23}  & -\Sigma_3 & \cdots & \bar{\Omega}_{n3}\\
\vdots & \vdots  & \vdots & \ddots & \vdots \\
\bar{\Omega}_{1n} & \bar{\Omega}_{2n} & \bar{\Omega}_{3n} & \cdots & -\Sigma_{n} \end{array} \right)\left(\begin{array}{c} p_1\\p_2\\p_3\\ \vdots \\p_n \end{array}\right)=0,
\end{eqnarray} 
i.e., as $\sum_j \rm{M}_{ij} p_j=0$, where $\rm{M}$ is a $n\times n$ matrix with entries
\begin{eqnarray}
\label{matrix}
\rm{M}_{ij}=\bar{\Omega}_{ji}-\delta_{ij}\Sigma_i.
\end{eqnarray}
It has been noted earlier that, due to the periodic graph structure, in addition to the internal edge, there can be multiple exterior edges connecting the same types of states in different periods. These additional rates are contained in $\bar{\Omega}_{ij}$ which denotes the sum of all transition rates from a state of type $i$ into a state of type $j$. In the Master equation this is considered in the summation $\{j\}$ over the set of adjacent sites which includes the exterior edges. Denoting the transition rate matrix including the normalization (e.g., in the first row) by $\tilde{\rm{M}}$, that is
\begin{eqnarray}
\tilde{\rm{M}}=\left(\begin{array}{ccccc}
1 & 1 & 1 & \cdots & 1\\
\bar{\Omega}_{12} & -\Sigma_2 & \bar{\Omega}_{32} & \cdots & \bar{\Omega}_{n2}\\
\bar{\Omega}_{13} & \bar{\Omega}_{23}  & -\Sigma_3 & \cdots & \bar{\Omega}_{n3}\\
\vdots & \vdots  & \vdots & \ddots & \vdots \\
\bar{\Omega}_{1n} & \bar{\Omega}_{2n} & \bar{\Omega}_{3n} & \cdots & -\Sigma_{n} \end{array} \right),
\end{eqnarray}
the Master equation including normalization reads
\begin{eqnarray}
\left(\begin{array}{ccccc}
1 & 1 & 1 & \cdots & 1\\
\bar{\Omega}_{12} & -\Sigma_2 & \bar{\Omega}_{32} & \cdots & \bar{\Omega}_{n2}\\
\bar{\Omega}_{13} & \bar{\Omega}_{23}  & -\Sigma_3 & \cdots & \bar{\Omega}_{n3}\\
\vdots & \vdots  & \vdots & \ddots & \vdots \\
\bar{\Omega}_{1n} & \bar{\Omega}_{2n} & \bar{\Omega}_{3n} & \cdots & -\Sigma_{n} \end{array} \right)
\left(\begin{array}{c} p_1\\p_2\\p_3\\ \vdots \\p_n \end{array}\right)=\left(\begin{array}{c} 1\\ 0 \\ 0 \\ \vdots \\ 0 \end{array}\right).
\end{eqnarray}
We therefore find the occupancies $p_i$ by matrix inversion as $\tilde{\rm{M}}^{-1}\mathbf{e}_1$, i.e. in the first column of the inverse matrix $\tilde{\rm{M}}^{-1}$.

A new relationship between the current $J$ and the transition rates is obtained when we take the derivative (denoted by a prime) of Eq.~(\ref{Qpi}) with respect to $\nu$:
\begin{eqnarray}
\label{flux1}
Q'=\sum_{\{j\}}(\Delta q'_{ji}+\Delta x_{ji})\Omega_{ij},
\end{eqnarray}
for every state $i$. Here and in the following the $\nu$-dependence is dropped. Using $\Delta q'_{ji}=q'_j-q'_i$ we can rewrite these equations as 
\begin{eqnarray}
\label{flux2}
Q'+q'_{i}\Sigma_i-\sum_{\{j\}}q'_{j}\Omega_{ij}=\sum_{\{j\}}\Delta x_{ji}\Omega_{ij}.
\end{eqnarray}
Since all relevant physical information is contained in the differences rather than the individual $q_i$'s, we have the freedom to fix one boundary value. If we choose $q_1=const$, obviously $q_1'=0$ and thus $q_i'=-\Delta q_{1i}'$. Defining the two column vectors $\mathbf{q}'\equiv(Q',\Delta q_{12}',...,\Delta q_{1n}')^T$ and $\mathbf{a}=(a_1,...,a_n)^T$ with $a_i\equiv\sum_{\{j\}}\Delta x_{ji}\Omega_{ij}$, we can cast the system of equations (\ref{flux2}) into the following matrix form
\begin{eqnarray}
\label{flux_Qmatrix}
\left(\begin{array}{ccccc}
1 & \bar{\Omega}_{12} & \bar{\Omega}_{13} & \cdots & \bar{\Omega}_{1n}\\
1 & -\Sigma_2 & \bar{\Omega}_{23} & \cdots & \bar{\Omega}_{2n}\\
1 & \bar{\Omega}_{32}  & -\Sigma_3 & \cdots & \bar{\Omega}_{3n}\\
\vdots & \vdots  & \vdots & \ddots & \vdots \\
1 & \bar{\Omega}_{n2} & \bar{\Omega}_{n3} & \cdots & -\Sigma_{n} \end{array}\right)
\left(\begin{array}{c} Q'\\\Delta q_{12}'\\\Delta q_{13}'\\ \vdots \\\Delta q'_{1n} \end{array}\right)
=\left(\begin{array}{c} a_1\\a_2\\a_3\\ \vdots \\a_n \end{array}\right).
\end{eqnarray}
We realize that the left hand side reveals the transpose matrix of $\tilde{\rm{M}}$. Eq.~(\ref{flux_Qmatrix}) is therefore equally expressed as
\begin{eqnarray}
\label{flux_matrix}
\tilde{\rm{M}}^{T}\mathbf{q}'=\mathbf{a}.
\end{eqnarray}
With this result we can rederive the fundamental relationship Eq.~(\ref{Qflux}) between the flux potential $Q$ and the flux $J$. Eq.~(\ref{flux_matrix}) is formally solved via matrix inversion
\begin{eqnarray}
\label{flux_matrix_sol}
\mathbf{q}'=\left(\tilde{\rm{M}}^T\right)^{-1}\mathbf{a}.
\end{eqnarray}
For any matrix the transpose of the inverse matrix is the inverse of the transpose matrix: $(\tilde{\rm{M}}^T)^{-1}=(\tilde{\rm{M}}^{-1})^T$. Since $\tilde{\rm{M}}^{-1}\mathbf{e}_1$ is the formal solution of the Master equation, we can deduce that the first row of $(\tilde{\rm{M}}^T)^{-1}$ contains the steady state occupancies $p_i$. Solving Eq.~(\ref{flux_matrix_sol}) for $Q'$ with the given expressions $a_i$ therefore leads to
\begin{eqnarray}
Q'=\sum_{i}p_i\sum_{\{j\}}\Delta x_{ji}\Omega_{ij},
\end{eqnarray}
so that by comparison with Eq.~(\ref{flux_def}), we can conclude that $Q'=J$, i.e., Eq.~(\ref{Qflux}) holds. We have therefore shown that
\begin{eqnarray}
\label{flux_int}
J=\sum_{\{j\}}(\Delta q'_{ji}+\Delta x_{ji})\Omega_{ij},
\end{eqnarray}
for every state $i$, which is a new representation of the steady state current $J$ in terms of $\Delta q'_{ji}$ instead of the microstate distributions $p_i$.

At this point it is instructive to associate a physical interpretation with the $\nu$ derivative of the $\Delta q_{ji}$. Taking the derivative of Eq.~(\ref{q_def}) with respect to $\nu$ leads to the following expression:
\begin{eqnarray}
\label{dq_prime}
\Delta q_{ji}'=\lim_{\tau\rightarrow\infty}\left[\frac{\int_{-\infty}^\infty \gamma\, p_\tau^{eq}(\gamma|j)\,e^{\nu\gamma}\,\upd\gamma}{\int_{-\infty}^\infty p_\tau^{eq}(\gamma|j)\,e^{\nu\gamma}\,\upd\gamma}-\frac{\int_{-\infty}^\infty \gamma\, p_\tau^{eq}(\gamma|i)\,e^{\nu\gamma}\,\upd\gamma}{\int_{-\infty}^\infty p_\tau^{eq}(\gamma|i)\,e^{\nu\gamma}\,\upd\gamma}\right].
\end{eqnarray}
In the discussion of the path-entropy maximization in Sec.~\ref{Sec_NCDB} we obtained the fundamental result (Eq.~(\ref{driven_p})) that in the driven ensemble the trajectories are reweighted with the exponential factor $e^{\nu\gamma}$. This implies that in the present context $p_\tau^{eq}(\gamma|j)\,e^{\nu\gamma}$ can be related to the unnormalized conditional probability that the system exhibits shear $\gamma$ from state $i$ \textit{under the driven dynamics}, i.e.,
\begin{eqnarray}
p_\tau^{dr}(\gamma|i)&\propto&p_\tau^{eq}(\gamma|i)e^{\nu\gamma},
\end{eqnarray}
in the limit of large $\tau$. Introducing
\begin{eqnarray}
\left<\gamma_i(\tau,\nu)\right>\equiv \frac{\int_{-\infty}^\infty \gamma\, p_\tau^{eq}(\gamma|i)\,e^{\nu\gamma}\,\upd\gamma}{\int_{-\infty}^\infty p_\tau^{eq}(\gamma|i)\,e^{\nu\gamma}\,\upd\gamma},
\end{eqnarray}
which denotes the \textit{mean} shear that the system accumulates over time $\tau$ in the steady state from state $i$, one can then express Eq.~(\ref{dq_prime}) as
\begin{eqnarray}
\Delta q_{ji}'=\lim_{\tau\rightarrow\infty}\left[\left<\gamma_j(\tau,\nu)\right>-\left<\gamma_i(\tau,\nu)\right>\right],
\end{eqnarray}
i.e., as the difference in mean shear between states $i$ and $j$ over infinite time. Furthermore, from Eq.~(\ref{flux_matrix_sol}) we find that the quantities $\Delta q_{ji}'$ can be determined using Cramer's rule \cite{EvansM02}:
\begin{eqnarray}
\label{qpr}
\Delta q_{1i}'=\frac{\det\left[(\tilde{\rm{M}}^T)^{(i)}\right]}{\det\left[\tilde{\rm{M}}^T\right]},
\end{eqnarray}
where $(\tilde{\rm{M}}^T)^{(i)}$ means that the $i$th column of $\tilde{\rm{M}}^T$ has to be replaced by $\mathbf{a}$. From the set of $\Delta q_{1i}'$ all other quantities $\Delta q_{ji}'$ follow by subtraction: $\Delta q_{ji}'=\Delta q_{1i}'-\Delta q_{1j}'$.

Relationship (\ref{flux_int}) is remarkable because the stationary shear current is determined from the properties of a \textit{single} state and its neighbors only, instead of the average Eq.~(\ref{flux_def}) over the whole basic graph. It does not involve energetics, since the knowledge of two constant `shear values' in addition to the transition rates is sufficient to determine the current. One constant, $\Delta x_{ji}$, measures the immediate difference in shear between states $i$ and $j$ and the other, $\Delta q'_{ji}$, the difference in mean shear that the system accumulates in the steady state over infinite time from states $i$ and $j$. The set of $\Delta q'_{ji}$ is directly related to the transition rate matrix $\tilde{\rm{M}}$ via Eq.~(\ref{qpr}).

From a computational point of view, no advantage is gained by determining the flux via Eq.~(\ref{flux_int}) if the quantities $\Delta q'_{ji}$ are determined by the same formal matrix inversion method as the probability distributions. However, Eq.~(\ref{flux_int}) provides us with a new interpretation of the steady state current and of the elements of the inverse transition rate matrix $\tilde{\rm{M}}^{-1}$. The quantities $\Delta q'_{ji}$ have a precise physical meaning independent of the NCDB formalism. It might be possible to identify related current expressions for other steady state systems that are not contained in the class of systems for which NCDB is valid.

\subsection{Fluctuation relation for the shear current of a trajectory}
\label{Sec_FT}

\textit{Fluctuation theorems} are mathematical relations for the fluctuations of thermodynamic quantities like heat (entropy production) or work in nonequilibrium systems (see \cite{Harris07} and references therein). In a nonequilibrium steady state the so called asymptotic or steady state fluctuation theorem (SSFT) states that the probability distribution $p_\tau(\rho)$ of finding a particular value of the thermodynamic quantity $\rho$ over time $\tau$ satisfies a certain symmetry relation of the form (cf. \cite{EvansD93,Gallavotti95,Kurchan98,Lebowitz99})
\begin{eqnarray}
\label{conventional}
\frac{p_\tau(\rho)}{p_\tau(-\rho)}\cong e^{c\rho\tau},
\end{eqnarray}
where $\cong$ indicates the asymptotic behavior for large $\tau$ and $c$ is a constant. The SSFT Eq.~(\ref{conventional}) (also refered to as Gallavotti-Cohen fluctuation theorem) represents a refinement of the second law of thermodynamics in that it quantifies the probability of observing temporary second law violations (negative $\rho$) in the nonequilibrium steady state. Relations similar to Eq.~(\ref{conventional}) have been derived for a variety of systems with different thermostatting mechanisms. For deterministic systems in a compact phase space the SSFT for the entropy production is expected to hold universally under the chaotic hypothesis \cite{Gallavotti95}. However, stochastic systems do not exhibit the same generality in their fluctuation behavior. Here, the validity of Eq.~(\ref{conventional}) relies, e.g., on the characteristics of the noise, or on the thermodynamic quantity considered. In a particular paradigmatic nonequilibrium particle model the SSFT has been shown to hold for the heat fluctuations, when the system is thermostatted by an equilibrium heat bath with Gaussian white noise characteristics \cite{VanZon03}. However, when one considers the heat fluctuations, or the work fluctuations under the influence of non-Gaussian noise, the SSFT is violated \cite{VanZon03b,Touchette07,Touchette09,Baule09,Baule09b}.

In a shear flow, a fluid region in the bulk of the volume receives noise from the sheared fluid surrounding it, which represents a nonequilibrium heat bath. NCDB quantifies the stochastic influence of such a nonequilibrium heat bath on the dynamics of the fluid. In systems satisfying NCDB a relation in the form of the SSFT holds for the fluctuations of the shear current of an individual trajectory and can be derived in a straightforward way. 
Our starting point is  the nonequilibrium path distribution $p^{dr}(\Gamma)$ of Eq.~(\ref{driven_p}). Using this path distribution, the distribution of the shear current $j$ of trajectories of duration $\tau$ can be formally expressed as
\begin{eqnarray}
p_\tau(j)=\frac{\sum_\Gamma \delta(j-\gamma(\Gamma)/\tau)p^{dr}(\Gamma)}{\sum_\Gamma p^{dr}(\Gamma)},
\end{eqnarray}
where the total shear of a phase space trajectory $\Gamma$ is denoted by $\gamma(\Gamma)$. Substituting Eq.~(\ref{driven_p}) yields
\begin{eqnarray}
p_\tau(j)\cong e^{\nu j \tau}\frac{\sum_\Gamma \delta(j-\gamma(\Gamma)/\tau)p^{eq}(\Gamma)}{\sum_\Gamma e^{\nu\gamma(\Gamma)}p^{eq}(\Gamma)}.
\end{eqnarray}
Here, the numerator is just the unnormalized probability distribution of observing the shear current $j$ at equilibrium. Clearly, this distribution has to be symmetric under a change of sign of $j$, i.e., at equilibrium the probability of observing a shear current $j$ over time $\tau$ is the same as observing $-j$ over the same time period. Taking the ratio $p_\tau(j)/p_\tau(-j)$ then immediately implies
\begin{eqnarray}
\label{NCDB_FT}
\frac{p_\tau(j)}{p_\tau(- j)}\cong e^{2\nu j\tau},
\end{eqnarray}
for large $\tau$. The distribution $p_\tau(j)$ depends implicitly on the parameter $\nu$.

The current fluctuation relation Eq.~(\ref{NCDB_FT}) characterizes the fluctuations of $j$ in the steady state and implies that the system's trajectory is overwhelmingly likely to follow the direction prescribed by the imposed driving, which is specified by the sign of the flux conjugated parameter $\nu$. Similar fluctuation relations for the current have previously been derived in the context of lattice gas models \cite{Derrida07} (see also \cite{Harris07} and references therein). In the present case we have shown the validity of such a current fluctuation relation for a very large class of nonequilibrium steady state systems, namely fluid flows under continuous shear, as described by NCDB. Eq.~(\ref{NCDB_FT}) follows then very naturally due to the Gibbs property of the path distribution $p^{dr}(\Gamma)$. The general relationship between the SSFT and Gibbs distributions has been discussed in \cite{Maes99}.

In the remainder of this article we apply the formal results of the previous sections to two simple Markovian jump models.

\section{Analytic solution of the two state graph}
\label{Sec_2}

As our first example we discuss a basic graph with two different types of states $1$ and $2$. For generality, an asymmetric separation between the states is assumed. Two distinct arrangements are then possible: (i) a simple connected path leading to a zig-zag shaped graph, and (ii) a graph in the form of connected triangles. (Other connectivities are also possible, one of which was considered in \cite{EvansR05}.) For both types of graphs the driven transition rates can be calculated exactly using the results of the previous sections. Obviously, (i) is a special case of (ii) where the horizontal transition rates are set to zero. For clarity we first discuss the simpler zig-zag graph in some detail in Secs.~\ref{Sec_2state_a} and \ref{Sec_2state_b} and then present results for case (ii) in Sec.~\ref{Sec_2state_c}. 

Let us first consider a zig-zag arrangment of the two states leading to a ratchet shaped state space. Although the solution of this two state model follows quite intuitively, we apply the network rules of Sec.~\ref{Sec_net} in a systematic way.

\subsection{Network rules}
\label{Sec_2state_a}

\textit{Edge rule}. The basic graph in Fig.~\ref{Fig_2zig} reveals one interior edge and one pair of exterior edges. In the following we use an intuitive notation for (u)pwards and (d)ownwards rates in positive (+) and negative direction (-). Driven rates are capitalized while their equilibrium counterparts are lower case. To the interior edge the transition rates $U^+\equiv\Omega_{12}$ and $D^-\equiv\Omega_{21}$ are associated, which depend on $\Delta q_{21}$ and carrying the shear $\pm\Delta x_{21}$. To the pair of exterior edges we associate the rates $U^-\equiv\Omega_{10}$ and $D^+\equiv\Omega_{23}$, which depend on $\Delta q_{32}$ and $\pm\Delta x_{32}$ (note that $\Delta q_{32}=\Delta q_{10}$ and $\Delta x_{32}=\Delta x_{10}$). At equilibrium the number of different transition rates remains four due to the asymmetry of the state space, but equilibrium detailed balance requires that $u^-/d^+=u^+/d^-=e^{-\beta\Delta E}$, with $\Delta E$ being the energy difference between states of type $1$ and $2$. Throughout the following calculations the inverse temperature $\beta$ is set to unity. According to Eq.~(\ref{NCDB_intro}) the driven transition rates are then given by:
\begin{eqnarray}
\label{twozigrates}
U^+&=&u^+e^{\nu\Delta x_{21}+\Delta q_{21}},\qquad D^-=d^-e^{-\nu\Delta x_{21}-\Delta q_{21}},\nonumber\\
U^-&=&u^-e^{-\nu\Delta x_{32}-\Delta q_{32}},\qquad D^+=d^+e^{\nu\Delta x_{32}+\Delta q_{32}}.
\end{eqnarray}

\begin{figure}
\begin{center}
\includegraphics[height=5cm]{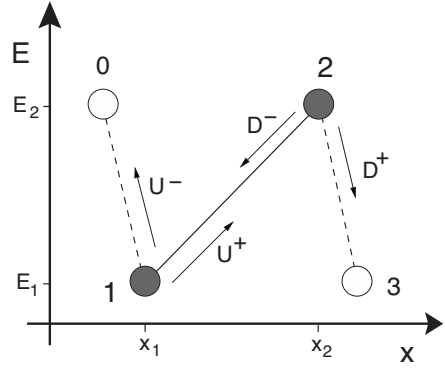}
\caption{\label{Fig_2zig}Basic graph of the two state zig-zag model. The dotted lines denote exterior edges.}
\end{center}
\end{figure}

\textit{Vertex rule.} There are two distinct vertices in the basic graph. To each corresponds an equation in the form of Eq.~(\ref{Qpi})
\begin{eqnarray}
Q&=&U^+ +U^- -(u^+ +u^-),\nonumber\\
Q&=&D^+ +D^- -(d^+ +d^-).
\end{eqnarray}

\textit{Loop rule.} There is one external closed loop $1\rightarrow 2\rightarrow 3$ corresponding to subsequent transitions $U^+$ and $D^+$. The loop constraint for the $\Delta q$'s then reads
\begin{eqnarray}
\Delta q_{21}+\Delta q_{32}=0.
\end{eqnarray}

In total we have the system of equations
\begin{eqnarray}
u^+e^{\nu\Delta x_{21}+\Delta q_{21}}+u^-e^{-\nu\Delta x_{32}-\Delta q_{32}}-(u^+ +u^-)&=&Q,\\
d^-e^{-\nu\Delta x_{21}-\Delta q_{21}}+d^+e^{\nu\Delta x_{32}+\Delta q_{32}}-(d^+ +d^-)&=&Q,\\
\Delta q_{21}+\Delta q_{32}&=&0,
\end{eqnarray}
for the three unknowns $Q$, $\Delta q_{21}$, and $\Delta q_{32}$. The solution can be found in a straightforward way. We simplify notation by setting $\Delta q\equiv \Delta q_{21}=-\Delta q_{32}$, as well as $\Delta x_1\equiv\Delta x_{21}$ and $\Delta x_2\equiv\Delta x_{32}$. Furthermore, we use the notation for the total exit rates $\sigma_1=u^++u^-$ and $\sigma_2=d^++d^-$. Substitution of the loop constraint yields
\begin{eqnarray}
(u^+e^{\nu\Delta x_{1}}+u^-e^{-\nu\Delta x_{2}})e^{\Delta q}-\sigma_1&=&Q,\label{2zig_eq1}\\
(d^-e^{-\nu\Delta x_{1}}+d^+e^{\nu\Delta x_{2}})e^{-\Delta q}-\sigma_2&=&Q.
\end{eqnarray}
Elimination of $e^{\Delta q}$ in this set of equations leads to a quadratic equation for $Q$, namely
\begin{eqnarray}
Q^2+(\sigma_1 +\sigma_2)Q+(\sigma_1\sigma_2-(u^+e^{\nu\Delta x_{1}}+u^-e^{-\nu\Delta x_{2}})(d^-e^{-\nu\Delta x_{1}}+d^+e^{\nu\Delta x_{2}}))=0.\nonumber\\
\end{eqnarray}
Due to equilibrium detailed balance we furthermore have
\begin{eqnarray}
\label{2loop_Qrel}
\sigma_1\sigma_2-(u^+e^{\nu\Delta x_{1}}+u^-e^{-\nu\Delta x_{2}})(d^-e^{-\nu\Delta x_{1}}+d^+e^{\nu\Delta x_{2}})=2u^+d^+(1-\cosh(\nu(\Delta x_1+\Delta x_2))).
\end{eqnarray}
The positive root then reads
\begin{eqnarray}
\label{twozig_fluxpot}
Q(\nu)&=&\frac{1}{2}\sqrt{(\sigma_1 +\sigma_2)^2 +8u^+d^+(\cosh(\nu(\Delta x_1+\Delta x_2))-1)}-\frac{1}{2}(\sigma_1 +\sigma_2).
\end{eqnarray}
In turn, $\Delta q$ is determined from Eq.~(\ref{2zig_eq1}) as
\begin{eqnarray}
\label{2zig_Dq}
\Delta q&=&\ln\left[\frac{Q+\sigma_1}{u^+e^{\nu\Delta x_1}+u^-e^{-\nu\Delta x_2}}\right].
\end{eqnarray}
For a given set of equilibrium rates, the driven transition rates are now completely determined as functions of the flux conjugated parameter $\nu$. With $Q$ and $\Delta q$ we can furthermore calculate the shear current $J(\nu)=\upd Q(\nu)/\upd\nu$ and the mean shear difference $\Delta q'=\upd \Delta q(\nu)/\upd \nu$ (see also below). Equilibrium is characterized by $\nu=0$. As expected the flux then vanishes: $J(\nu=0)=0$. One can also easily see that for $\nu=0$ both $Q(0)=0$ and $\Delta q(0)=0$, and the driven rates Eqs.~(\ref{twozigrates}) reduce to the equilibrium rates. 

\subsection{Invariant quantities}
\label{Sec_2state_b}

Since the asymmetric two state zig-zag graph is obviously a simple connected path, we can alternatively determine the driven transition rates using the $\nu$-independent representation. To this end we have to set up the invariant quantities and find the relation between the rates and the particle flux.
The invariant quantities follow immediately from Eqs.~(\ref{pcons}) and (\ref{qcons}):
\begin{eqnarray}
\label{eqset1}
U^+\,D^-&=&u^+d^-,\\
D^+\,U^-&=&d^+u^-,\\
\label{eqset2}
D^++D^--(U^++U^-)&=& d^+ +d^- -(u^+ +u^-).
\end{eqnarray}

Finally, to find the flux, we calculate the probability distributions of states $1$ and $2$. The Master equation for the two state graph has the following matrix form
\begin{eqnarray}
\left(\begin{array}{cc}
-(U^++U^-) & D^++D^- \\
U^++U^- & -(D^++D^-)
\end{array} \right)\left(\begin{array}{cc} p_1 \\ p_2 \end{array}\right)=0.
\end{eqnarray}
Including the normalization in the transition matrix leads to
\begin{eqnarray}
\tilde{\rm{M}}\left(\begin{array}{cc} p_1 \\ p_2 \end{array}\right)=\left(\begin{array}{cc} 1 \\ 0 \end{array}\right),
\end{eqnarray}
where
\begin{eqnarray}
\label{2_invM}
\tilde{\rm{M}}=\left(\begin{array}{cc}
 1 & 1 \\
U^++U^- & -(D^++D^-)
\end{array} \right).
\end{eqnarray}
Matrix inversion then yields the probability distributions (or occupancies) $p_1$, $p_1$, which, in turn allow for the calculation of the current $J$ via Eq.~(\ref{flux_def}). The results for the probability distributions as functions of the driven transition rates are
\begin{eqnarray}
p_1&=&\frac{D^++D^-}{U^++U^-+D^++D^-},\label{2zig_occup}\\
p_2&=&\frac{U^++U^-}{U^++U^-+D^++D^-}.\label{2zig_occup2}
\end{eqnarray}

According to the results of Sec.~\ref{Sec_cur} we can alternatively determine the current and the mean shear differences $\Delta q'$ from Eq.~(\ref{flux_matrix_sol}). This equation is given by
\begin{eqnarray}
\left(\begin{array}{cc} J \\ -\Delta q' \end{array}\right)=\left(\tilde{\rm{M}}^{-1}\right)^T\left(\begin{array}{cc} U^+\Delta x_1 -U^- \Delta x_2 \\ D^+\Delta x_2-D^- \Delta x_1 \end{array}\right).
\end{eqnarray} 
The matrix algebra is straightforward and yields
\begin{eqnarray}
\label{twozig_flux2}
J&=&(\Delta x_1+\Delta x_2)\frac{U^+D^+-U^-D^-}{U^++U^-+D^++D^-},\\
\Delta q'&=&\frac{(D^+ +U^-)\Delta x_2 -(U^+ +D^-)\Delta x_1}{U^++U^-+D^++D^-}.
\end{eqnarray}
The last two expressions agree with the ones obtained in the $\nu$-representation, if the driven rates are substituted according to Eq.~(\ref{twozigrates}) with the known expressions for $Q$ and $\Delta q$, Eqs.~(\ref{twozig_fluxpot}) and~(\ref{2zig_Dq}).

For a given set of equilibrium rates and a prescribed particle flux, the transition rates in the driven steady state are determined as solutions of the set of equations~(\ref{eqset1})---(\ref{eqset2}) and (\ref{twozig_flux2}). In this case the driven rates are parametrized by $J$.

\subsection{Two state graph with a loop}
\label{Sec_2state_c}

If we add two additional external edges $2\rightarrow 0$ and $2\rightarrow 4$ to the two state zig-zag graph of Fig.~\ref{Fig_2zig}, we obtain a state space with the structure of Fig.~\ref{Fig_2loop}. The system in state $2$ can now choose to reach state $4$ directly or by going first `downhill' to a state of type $1$ and then `uphill'. The choice will depend on which path is more favourable to achieve the imposed flux $J$ for given parameter values. As before, NCDB precisely quantifies the change in the transition rates under driving.

\begin{figure}
\begin{center}
\includegraphics[height=5cm]{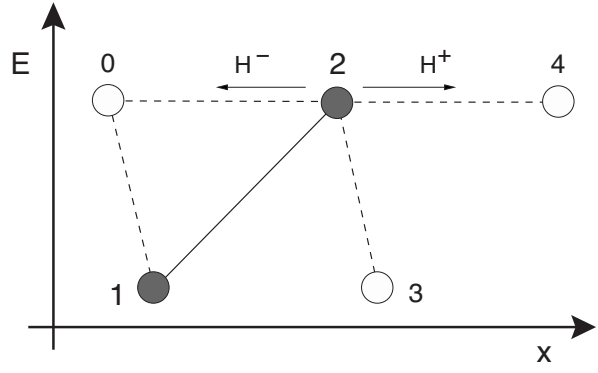}
\caption{\label{Fig_2loop}Basic graph of the two state loop model. The dotted lines denote exterior edges.}
\end{center}
\end{figure}

We denote the transition rates associated to the additional exterior edges with $H^+\equiv\Omega_{24}$ and $H^-\equiv\Omega_{20}$. Both rates are independent of $\Delta q$ since they connect states of the same type and symmetry requires that the corresponding equilibrium rates are equal. The rates are therefore fully specified as
\begin{eqnarray}
\label{2loop_h}
H^+=h\,e^{\nu(\Delta x_1+\Delta x_2)},\qquad H^+=h\,e^{-\nu(\Delta x_1+\Delta x_2)}.
\end{eqnarray}
The effect of the additional edges on the other rates can be calculated analogously to the zig-zag graph. The total exit rate relation for state $2$ (used in the {\em vertex rule}) has to be extended by the new rates, whereas the relation for state $1$ remains unchanged. This leads to
\begin{eqnarray}
Q&=&U^+ +U^- -(u^+ +u^-),\\
Q&=&D^++D^-+H^++H^--(d^++d^-+2h).
\end{eqnarray}
Introducing the total exit rate $\sigma_2=d^++d^-+2h$ and substituting the expressions for the driven rates, we can write the system of equations as
\begin{eqnarray}
(d^-e^{-\nu\Delta x_{1}}+d^+e^{\nu\Delta x_{2}})e^{-\Delta q}+2h\cosh(\nu(\Delta x_1+\Delta x_2)-\sigma_2&=&Q,\\
(u^+e^{\nu\Delta x_{1}}+u^-e^{-\nu\Delta x_{2}})e^{\Delta q}-\sigma_1&=&Q.
\end{eqnarray}
The quadratic equation for the flux potential then reads
\begin{eqnarray}
0&=&Q^2+(\sigma_1 +\sigma_2-2h\cosh(\nu(\Delta x_1+\Delta x_2))Q+\sigma_1\sigma_2-\sigma_12h\cosh(\nu(\Delta x_1+\Delta x_2)\nonumber\\
&&-(u^+e^{\nu\Delta x_{1}}+u^-e^{-\nu\Delta x_{2}})(d^-e^{-\nu\Delta x_{1}}+d^+e^{\nu\Delta x_{2}}),
\end{eqnarray}
with the positive root
\begin{eqnarray}
\label{2loop_Q}
Q(\nu)&=&\frac{1}{2}\sqrt{(\sigma_1+\sigma_2-2h\cosh(\nu(\Delta x_1+\Delta x_2)))^2+8(\sigma_1 h+u^+d^+)(\cosh(\nu(\Delta x_1+\Delta x_2))-1)}\nonumber\\
&&-\frac{1}{2}(\sigma_1+\sigma_2-2h\cosh(\nu(\Delta x_1+\Delta x_2))).
\end{eqnarray}
As above, $\Delta q$ is determined from Eq.~(\ref{2zig_Dq}).

Having thus specified all the driven rates, we can determine the remaining quantities in the matrix formalism. Since the rates $H^+$ and $H^-$ connect two states of the same type, their effect on the occupancies balances to zero. The transition rate matrix $\tilde{\rm{M}}$ is therefore identical to Eq.~(\ref{2_invM}) and the probability distributions are given by Eqs.~(\ref{2zig_occup}) and~(\ref{2zig_occup2}) as in the zig-zag case. The shear current $J$ and the mean shear difference $\Delta q'$ are determined by Eq.~(\ref{flux_matrix_sol}), which reads here
\begin{eqnarray}
\left(\begin{array}{cc} J \\ -\Delta q' \end{array}\right)=\left(\tilde{\rm{M}}^{-1}\right)^T\left(\begin{array}{cc} U^+\Delta x_1 -U^- \Delta x_2 \\ (D^++H^+-H^-) \Delta x_2-(D^-+H^--H^+) \Delta x_1 \end{array}\right).
\end{eqnarray} 
We thus obtain
\begin{eqnarray}
\label{2state_J}
J&=&(\Delta x_1+\Delta x_2)\frac{U^+D^+-U^-D^-+(U^++U^-)(H^+-H^-)}{D^+ +D^-+U^+ +U^-},\\
\Delta q'&=&\frac{(D^+ +U^-)\Delta x_2 -(U^+ +D^-)\Delta x_1+(H^+-H^-)(\Delta x_1+\Delta x_2)}{D^+ +D^-+U^+ +U^-}.
\end{eqnarray}

\begin{figure}
\begin{center}
\begin{tabular}{ll}
(a) \includegraphics[width=7.37cm]{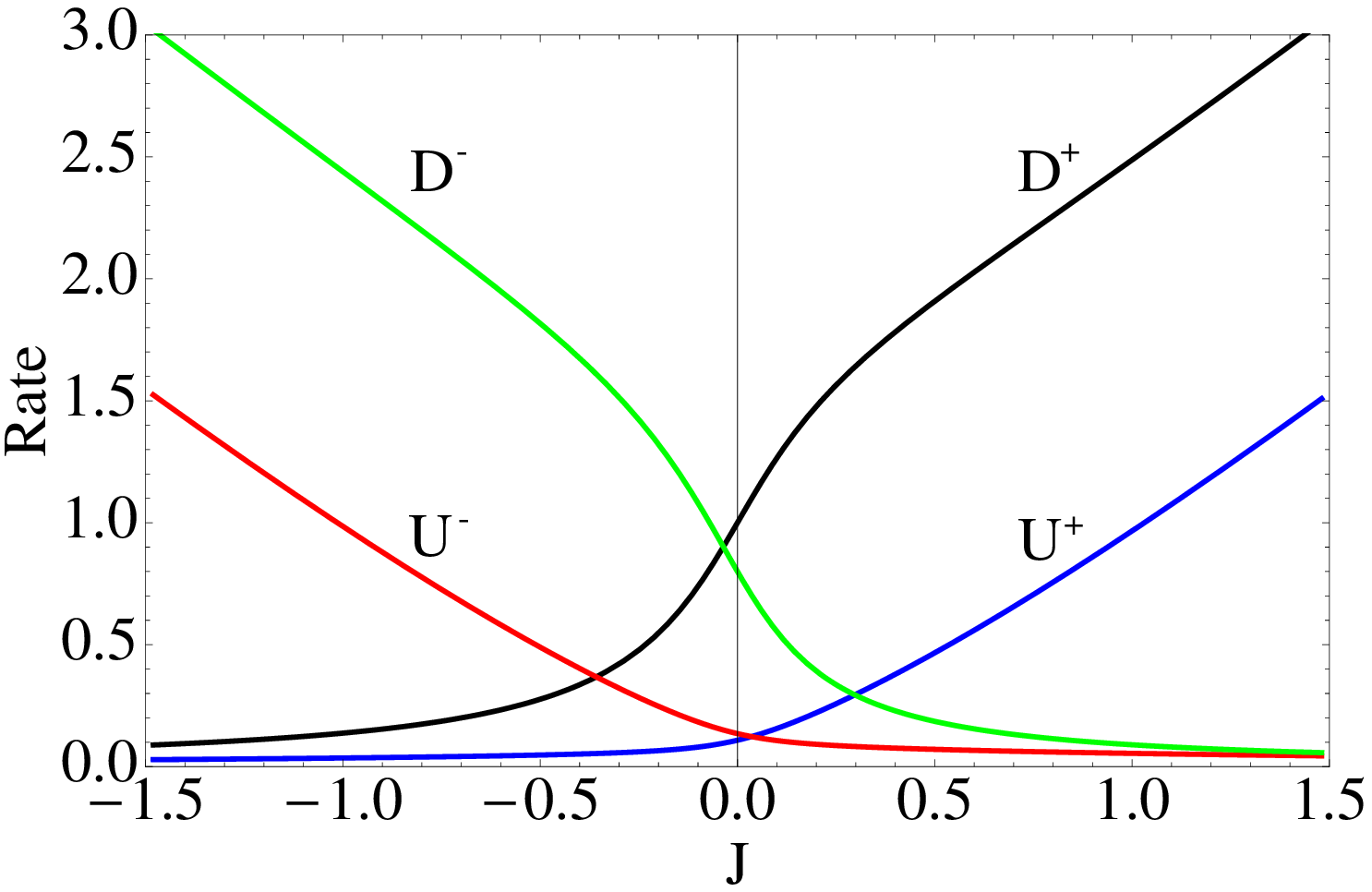} & (b) \includegraphics[width=7cm]{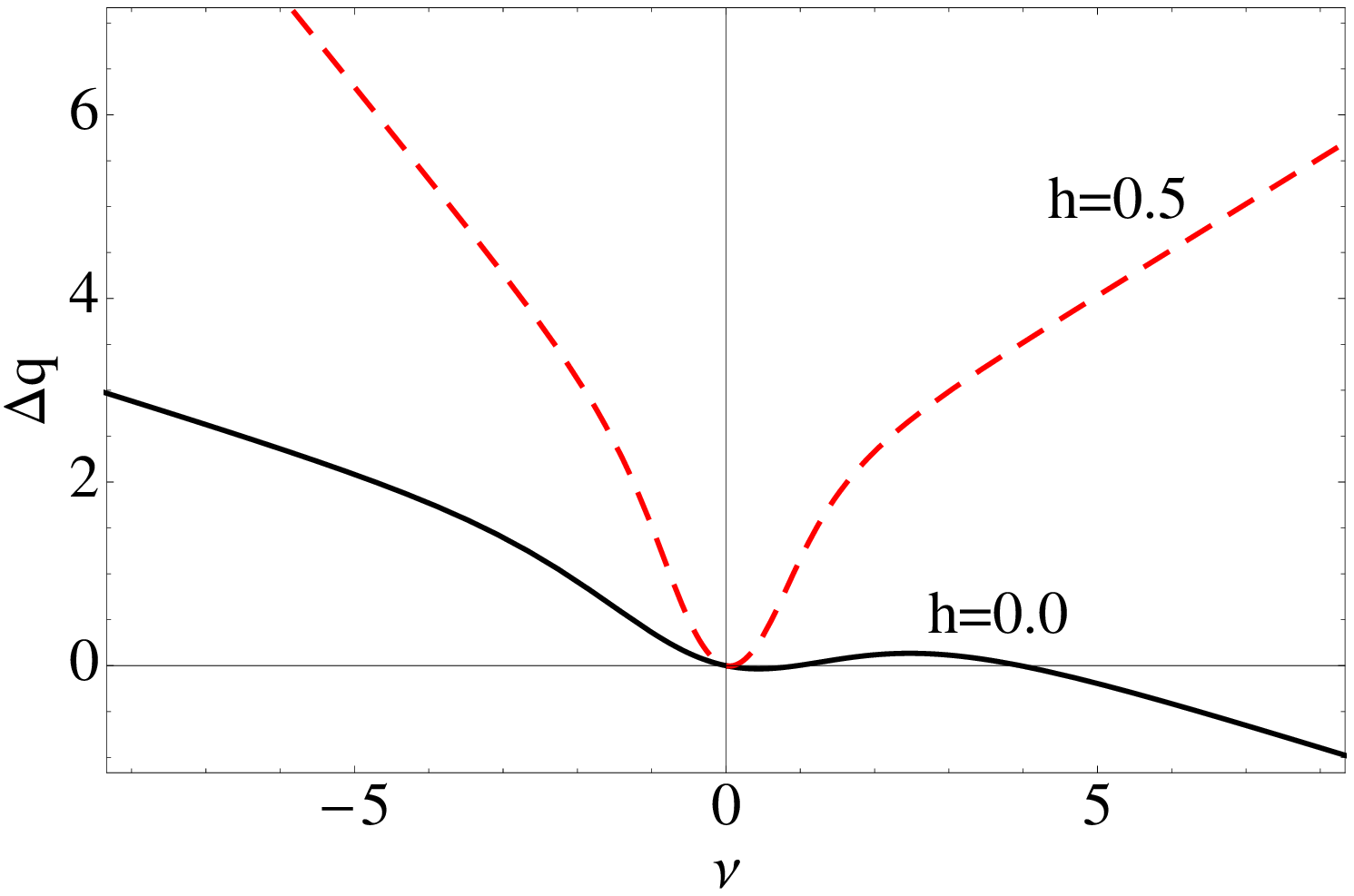}
\end{tabular}
\caption{\label{Fig_2zig}(a) The four transition rates of the two-state zig-zag model (h=0.0) plotted as functions of the current $J$. (b) The function $\Delta q(\nu)$ of Eq.~(\ref{2zig_Dq}) (with $Q(\nu)$ given by Eq.~(\ref{2loop_Q})) for two values of the equilibrium rate $h$. Parameter values:  $\Delta E=2.0$, $d^+=1.0$, $d^-=0.8$, $\Delta x_1=1.0$, $\Delta x_2=0.5$.}
\end{center}
\end{figure}

\begin{figure}
\begin{center}
\begin{tabular}{ll}
(a) \includegraphics[width=7.0cm]{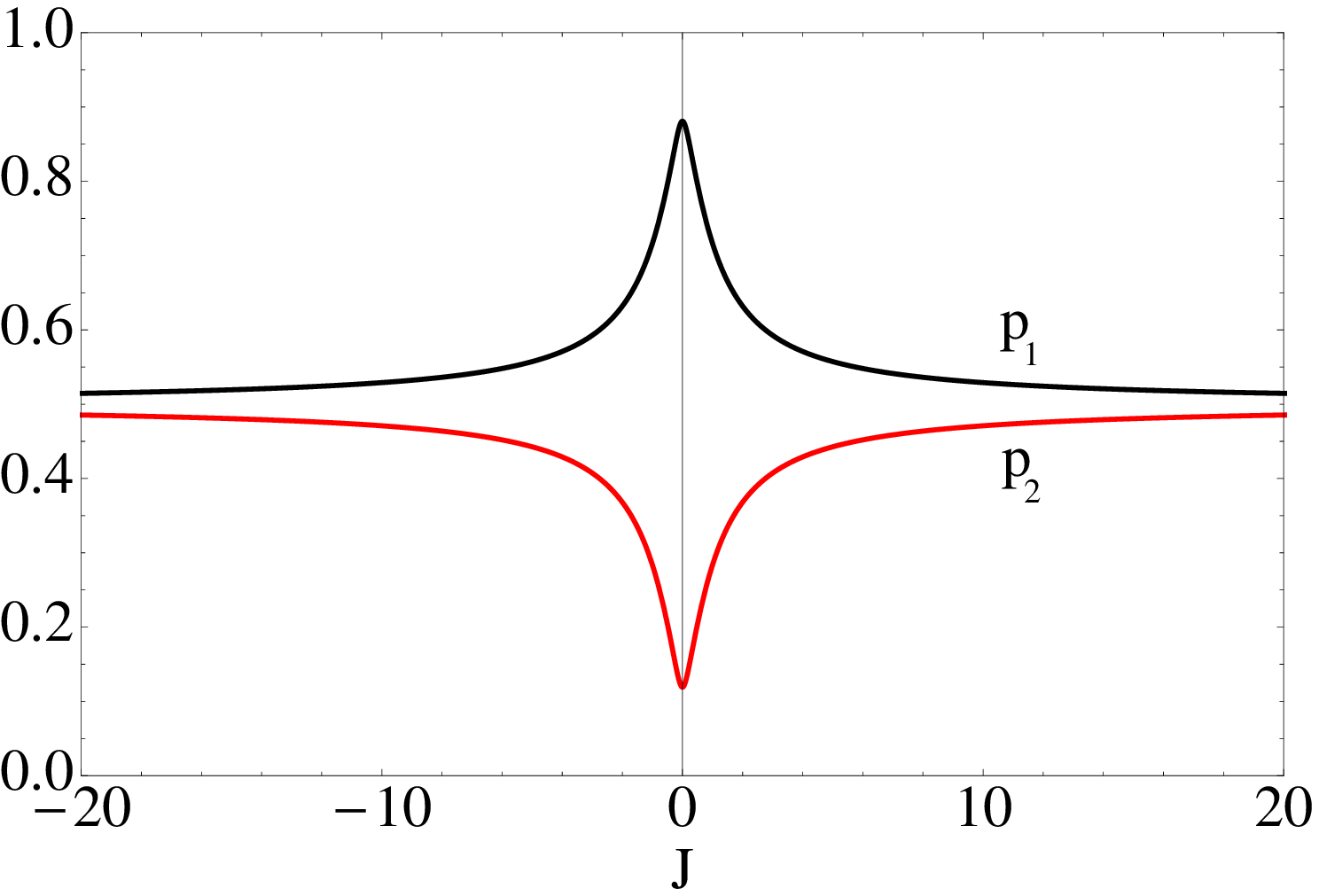} & (b) \includegraphics[width=7.0cm]{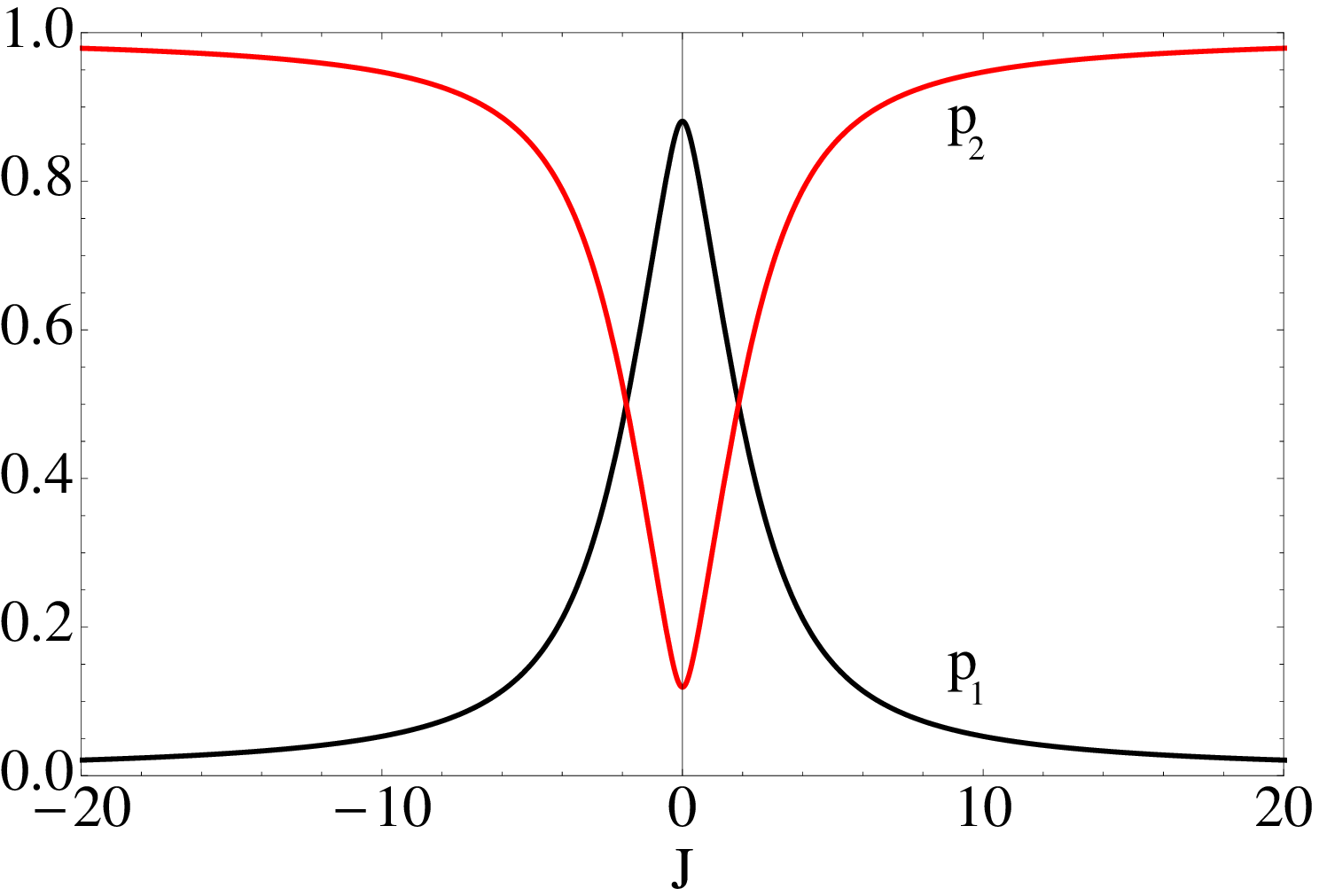}
\end{tabular}
\caption{\label{Fig_2state_p}The occupancies $p_1$ and $p_2$ of Eqs.~(\ref{2zig_occup}) and (\ref{2zig_occup2}) plotted as functions of the current $J$ for parameter values $\Delta E=2.0$, $d^+=1.0$, $d^-=0.8$, $\Delta x_1=1.0$, $\Delta x_2=0.5$. (a) In the two-state zig-zag model both distributions converge to $1/2$ for large $J$. (b) In the two-state loop model, with $h=0.5$. The occupancy of type 2 states, $p_2$ converges to $1$ for large $J$, while $p_1$ decays to zero.}
\end{center}
\end{figure}

\begin{figure}
\begin{center}
\includegraphics[width=10.0cm]{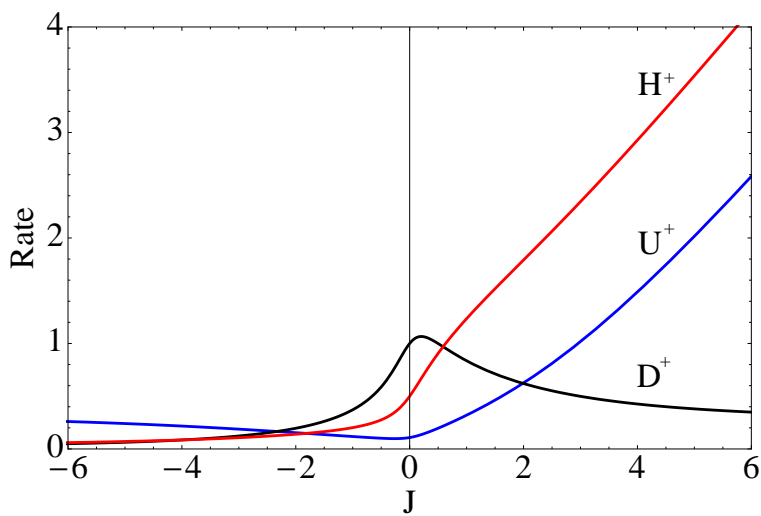}
\caption{\label{Fig_2loop}The three forward transition rates in the two-state loop model. Parameter values: $\Delta E=2.0$, $d^+=1.0$, $d^-=0.8$, $\Delta x_1=1.0$, $\Delta x_2=0.5$, $h=0.5$.}
\end{center}
\end{figure}

In the limit $h\rightarrow 0$ the two-state loop model obviously reduces to the zig-zag model discussed in the previous section. The four transition rates $U^+$, $U^-$, $D^+$, $D^-$ in this case are plotted in Fig.~\ref{Fig_2zig}(a) parametrically as functions of the current $J$ for given values of the parameters $\Delta E$, $d^+$, $d^-$, $\Delta x_1$, and $\Delta x_2$. Clearly, the transition rates in the direction of $J$ are enhanced for larger $J$, while the transition rates in the opposite direction are attenuated. For large $J$ the transition rates in the direction of the driving become proportional to $J$, as implied by the current relation Eq.~(\ref{2state_J}). The occupancies $p_1$, $p_2$, which at equilibrium ($J=0$) are determined by Boltzmann's law, converge to the value $1/2$ in the limit of $J\rightarrow \infty$, indicating that the energy difference between states $1$ and $2$ becomes irrelevant for strong driving (see Fig.~\ref{Fig_2state_p}(a)).

When $h\neq 0$ the system is able to gain a shear increment $\Delta x_1+\Delta x_2$ by going directly from state $2$ to state $4$, i.e., another state of type $2$, without first going to state $3$, a state of type $1$. This implies that, under (strong) forward driving, it is advantageous for the system in state $2$ to make the transition $2\rightarrow 4$ instead of the transition $2\rightarrow 3$, because it is then able to gain an increment $\Delta x_1+\Delta x_2$ in one step instead of two. This intuitively expected behavior of the two-state loop model is evident in Fig.~\ref{Fig_2loop}, where we plot the three forward transition rates under forward driving. While in the zig-zag case the rate $D^+$ is enhanced under increased forward driving, it is attenuated in the loop case due to the presence of the additional edge. The system finds that being in state $1$ is less favorable in order to achieve shear, than being in state $2$.

The quantity $\Delta q=\Delta q_{21}$, Eq.~(\ref{2zig_Dq}) (with $Q$ of Eq.~(\ref{2loop_Q})), is plotted in Fig.~\ref{Fig_2zig}(b) for the two cases $h=0$ and $h\neq 0$. In the zig-zag case ($h=0$) one notices that, for the given choice of the parameters $\Delta x_1>\Delta x_2$, $\Delta q$ becomes negative for large forward driving (large $\nu$), indicating that state $1$ has a larger propensity for future shear than state $2$. This is because from state $1$ the system can gain a larger (forward) shear increment ($\Delta x_1$) than from state $2$ ($\Delta x_2$). For weak driving (small $\nu$) a crossover behavior is observed, where $\Delta q$ exhibits local maxima and minima. In the loop case ($h\neq 0$) state $2$ is favored due to the additional edge and thus $\Delta q$ increases for both larger forward and backward driving. In fact, from the analytical expression Eq.~(\ref{2zig_Dq}) one finds that $\Delta q(\nu)\propto |\nu|$ for large $\pm\nu$ in the case $h\neq 0$, while $\Delta q(\nu)\propto \nu(\Delta x_2-\Delta x_1)$ in the case $h=0$.

\section{Three state graph}
\label{Sec_3}

\begin{figure}
\begin{center}
\includegraphics[height=5cm]{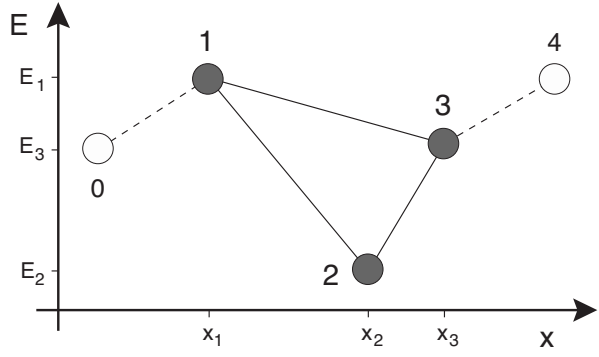}
\caption{\label{Fig_3state}Basic graph of a three state model with an interior loop. The dotted lines denote exterior edges.}
\end{center}
\end{figure}

A basic graph with three distinct states allows for a variety of periodic graph structures with different connectivities. In the following we consider a basic graph that contains an internal closed loop (see Fig.~\ref{Fig_3state}). For this arrangement the driven transition rates can only be found using the network rules, since the number of independent rates exceeds the number of invariant quantities plus current relation.

\textit{Edge rule.} There are three internal edges and one pair of external edges in the basic graph. The three internal edges correspond to six distinct rates: $\Omega_{12}$, $\Omega_{13}$, $\Omega_{23}$, and the associated reverse rates. The pair of external edges is associated with $\Omega_{10}$ and $\Omega_{34}$, where state $0$ is of the same type as $3$ and $4$ of the same type as $1$. The set of driven transition rates is specified as in Eq.~(\ref{NCDB_intro})
\begin{eqnarray}
\Omega_{ij}&=&\omega_{ij}\,e^{\nu\Delta x_{ji}+\Delta q_{ji}}.\nonumber
\end{eqnarray}
\textit{Vertex rule} (Eq.~(\ref{Qpi})) gives one relation for each vertex. These three equations read
\begin{eqnarray}
Q&=&\Omega_{12}+\Omega_{13}+\Omega_{10} -(\omega_{12}+\omega_{13}+\omega_{10}),\\
Q&=&\Omega_{23}+\Omega_{21} -(\omega_{23}+\omega_{21}),\\
Q&=&\Omega_{31}+\Omega_{32}+\Omega_{34}-(\omega_{31}+\omega_{32}+\omega_{34}).
\end{eqnarray}
\textit{Loop rule.} There is one external loop $0\rightarrow 1\rightarrow 3$ and one internal loop $1 \rightarrow 2 \rightarrow 3 \rightarrow 1$ so constraints on the four $\Delta q$'s read
\begin{eqnarray}
\Delta q_{10}+\Delta q_{31}&=&0,\\
\Delta q_{21}+\Delta q_{32}+\Delta q_{13}&=&0.
\end{eqnarray}
Furthermore the internal loop requires that $\Delta x_{31}=\Delta x_{21}+\Delta x_{32}$. We use the following simplifying notation:
\begin{eqnarray}
\ln z&\equiv&\Delta q_{21} ,\\
\ln y&\equiv&\Delta q_{32} ,\\
W_{ij}&\equiv& \omega_{ij}\,e^{\nu\Delta x_{ji}}.
\end{eqnarray}
The loop constraints then imply that $\Delta q_{01}=\Delta q_{31}=\ln(yz)$. In this notation the vertex rules read
\begin{eqnarray}
W_{12}\,z+ (W_{10}+W_{13})yz-\sigma_1&=&Q,\\
W_{23}\,y+ W_{21}\,z^{-1}-\sigma_2&=&Q,\\
(W_{31}+W_{34})(yz)^{-1}+ W_{32}\,y^{-1}-\sigma_3&=&Q.
\end{eqnarray}
where the total exit rates are as before $\sigma_i=\sum_{j}\omega_{ij}$. Elimination of $Q$ then leads to two quadratic equations
\begin{eqnarray}
\label{3loop_eq}
(W_{10}+W_{13})z^2y+W_{12}\,z^2-W_{23}\,yz+(\sigma_2-\sigma_1)z-W_{21}&=&0,\label{three_eq1}\\
W_{23}\,y^2z+(\sigma_3-\sigma_2)yz+W_{21}\,y-W_{32}\,z-(W_{31}+W_{34})&=&0.\label{three_eq2}
\end{eqnarray}
One realizes that already for the three state model an analytical solution is exceedingly difficult to obtain. Solving for example Eq.~(\ref{three_eq1}) for $y$ and substituting into Eq.~(\ref{three_eq2}) yields an equation for $z$ which is of fifth order and therefore not exactly solvable. Alternatively one can determine the driven transition rates in a straightforward way by solving Eqs.~(\ref{three_eq1}) and (\ref{three_eq2}) numerically for $y$ and $z$.

The current and probability distributions are easiest obtained from the matrix methods outlined in Sec.~\ref{Sec_cur}. For the three state graph of Fig.~\ref{Fig_3state} the master equation reads in matrix form
\begin{eqnarray}
\left(\begin{array}{ccc}
-(\Omega_{13}+\Omega_{12}+\Omega_{10}) & \Omega_{21} & \Omega_{31}+\Omega_{34} \\
\Omega_{12} & -(\Omega_{23}+\Omega_{21}) & \Omega_{32} \\
\Omega_{13}+\Omega_{10} & \Omega_{23} & -(\Omega_{31}+\Omega_{32}+\Omega_{34})
\end{array} \right)\left(\begin{array}{ccc} p_1 \\ p_2 \\ p_3 \end{array}\right)=0.
\end{eqnarray}
The probability distributions are then obtained as
\begin{eqnarray}
\label{3loop_p}
\left(\begin{array}{ccc} p_1 \\ p_2 \\ p_3 \end{array}\right)=\tilde{\rm{M}}^{-1}\left(\begin{array}{ccc} 1 \\ 0 \\ 0 \end{array}\right),
\end{eqnarray}
where
\begin{eqnarray}
\tilde{\rm{M}}=\left(\begin{array}{ccc}
1 & 1 & 1 \\
\Omega_{12} & -(\Omega_{23}+\Omega_{21}) & \Omega_{32} \\
\Omega_{13}+\Omega_{10} & \Omega_{23} & -(\Omega_{31}+\Omega_{32}+\Omega_{34})
\end{array} \right).
\end{eqnarray}
Using Eq.~(\ref{flux_matrix_sol}) we can then determine the shear current and the mean shear differences:
\begin{eqnarray}
\label{3loop_flux}
\left(\begin{array}{ccc} J \\ \Delta q'_{12} \\ \Delta q'_{13} \end{array}\right)=\left(\tilde{\rm{M}}^{-1}\right)^{T}\left(\begin{array}{ccc} \Omega_{13}(\Delta x_{32}+\Delta x_{21})+\Omega_{12}\Delta x_{21}-\Omega_{10}\Delta x_{43} \\ \Omega_{23}\Delta x_{32} - \Omega_{21}\Delta x_{21} \\ \Omega_{34}\Delta x_{43}-\Omega_{32}\Delta x_{32}-\Omega_{31}\Delta x_{31} \end{array}\right).
\end{eqnarray}

By solving Eqs.~(\ref{three_eq1}) and (\ref{three_eq2}) numerically for $\Delta q_{21}$ and $\Delta q_{32}$ given a set of equilibrium rates $\omega_{ij}$ and distances $\Delta x_{ji}$, one can determine the driven transition rates as well as the flux potential for the three state graph of Fig.~\ref{Fig_3state}. Eqs.~(\ref{3loop_p}) and (\ref{3loop_flux}) can then be used to obtain the probability distributions, the current and the mean shear differences.

\begin{figure}
\begin{center}
\includegraphics[width=10.0cm]{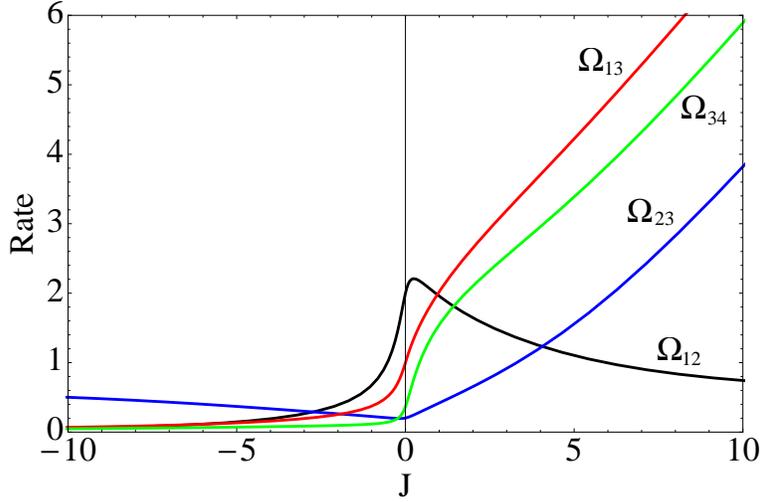}
\caption{\label{3state_rates}The four forward transition rates of the three-state loop model of Fig.~\ref{Fig_3state}. Parameter values: $\Delta E_{12}=3.0$, $\Delta E_{32}=2.0$, $\omega_{10}=1.0$, $\omega_{13}=1.0$, $\omega_{32}=1.5$, $\Delta x_{21}=1.8$, $\Delta x_{32}=1.0$, $\Delta x_{43}=0.5$.}
\end{center}
\end{figure}

The results for the driven transition rates in the forward direction are shown in Fig.~\ref{3state_rates}. For forward shear three of the transition rates are strongly enhanced. However, we observe that the transition $1\rightarrow 2$ is attenuated at high shear rate, even though it contributes a positive shear increment $\Delta x_{21}$. This indicates that the system disfavors the path via state $2$ that requires two transitions to acquire the shear increment $\Delta x_{32}$. For large shear rates the system will thus predominantly choose the high-mobility ``channel" $1\rightarrow 3$. A similar observation is made for the transition $2\rightarrow 3$, which, for backward driving, remains significant because it connects to the favorable direct channel.

\begin{figure}
\begin{center}
\includegraphics[width=7.3cm]{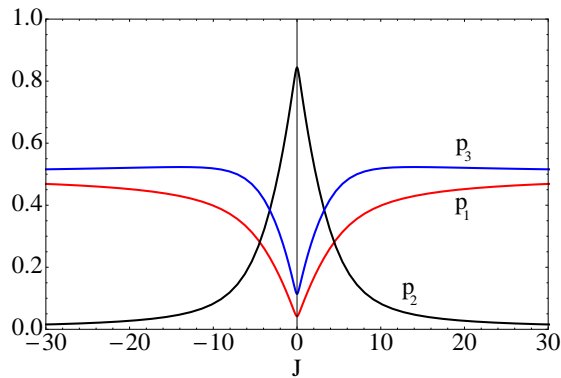}
\caption{\label{3state_p}The steady state distributions $p_1$, $p_2$, $p_3$ of Eq.~(\ref{3loop_p}) plotted parametrically as functions of the current $J$. Parameter values: $\Delta E_{12}=3.0$, $\Delta E_{32}=2.0$, $\omega_{10}=1.0$, $\omega_{13}=1.0$, $\omega_{32}=1.5$, $\Delta x_{21}=1.8$, $\Delta x_{32}=1.0$, $\Delta x_{43}=0.5$.}
\end{center}
\end{figure}

Accordingly, the steady state occupancies $p_1$ and $p_3$ converge to the value $1/2$ for strong driving in both forward and backward direction (cf. Fig.~\ref{3state_p}). On the other hand, the occupancy of state $2$, energetically favored over states $1$ and $3$ at equilibrium, decays to zero for larger shear.

\section{Summary}

The search for fundamental principles governing the behavior of systems in out of equilibrium situations has long been an area of intensive research. We have discussed a statistical theory (NCDB) that starts from first principles and governs the steady-state motion of any flowing system on which work is done by a \textit{weakly-coupled} nonequilibrium reservoir that is ergodic and microscopically reversible, such as a complex fluid under continuous shear. NCDB provides a description of the microscopic dynamics of these systems in the form of exact constraints on the transition rates in the driven steady state, arbitrarily far away from equilibrium, akin to the principle of detailed balance for equilibrium systems. In this article we have investigated this theory for systems evolving in discrete state spaces. We derived a simple relationship between the flux potential and the total exit rates, which leads to a number of important further results.

Two simple sets of invariant quantities have been formulated: (i) the product of forward and reverse transition rates and (ii) the differences in total exit rates, for every pair of microstates, equal the corresponding equilibrium values and are thus unchanged by the driving. These invariant quantities are non-trivial and experimentally accessible predictions of NCDB. We have devised a systematic method to determine the driven transition rates by setting up a system of equations from simple network rules. Furthermore, we have investigated properties of the shear current in systems satisfying NCDB. We show that the stationary shear current can be determined independently of the steady state distribution of microstates. Considering the shear current exhibited by an individual system trajectory, we show that the fluctuations of this shear current satisfy a fluctuation relation of the Gallavotti-Cohen type. The validity of such a fluctuation relation is related to the Gibbs form of the path distribution $p^{dr}(\Gamma)$ (cf. \cite{Maes99}) 

In a system with two types of states all quantities of the NCDB formalism can be expressed analytically. However, even for just three states results are only obtainable numerically. Both types of system show consistent behavior. In state-spaces containing a loop, the particular path favoured at highest driving is the one most accommodating for carrying flux.
Even in these simple models the non-local nature of NCDB becomes evident. Whereas mean-field theories without the quantity $\Delta q$ would simply boost any transition in the forward flux direction, NCDB takes into account the future propensity for achieving flux. That propensity depends on the global structure of the state space, and is communicated in the noise from the reservoir of other systems exploring the possibilities of the steady-state dynamics. A particular transition will thus be attenuated if it connects to a state which is blocked or from where subsequent transitions carry low flux. This striking property indicates that NCDB might ultimately be able to describe the counter-intuitive phase behavior exhibited for example by real complex fluids under shear.

Future work in this context will predominantly focus on further applications of NCDB to more realistic models as well as independent experimental tests. In particular the invariant quantities provide a straightforward criterion to check the validity of the theory (cf. \cite{nextPRL}). Even though these new exact relations are based on a rigorous statistical mechanical derivation, only comparison with experimental data can shed light on their significance for our understanding of nonequilibrium phenomena.

\acknowledgments

We are grateful to Hugo Touchette, Richard Blythe, Mike Cates, Rob Jack, and Peter Sollich for helpful discussions. The work was funded by EPSRC Grant GR/T24593/01 and by The Royal Society.


\end{document}